\documentclass[preprint,5p,times,twocolumn]{elsarticle}

\usepackage{amssymb}
\usepackage{amsmath}

\usepackage{url}

\journal{Computer Physics Communications}

\begin{document}

\begin{frontmatter}



\title{$\gamma$-Cascade V4: A Semi-Analytical Code for\\Modeling Cosmological Gamma-Ray Propagation}

\author[1]{Antonio Capanema}
\ead{antoniocggalvao@gmail.com}
\author[2,3]{Carlos Blanco}
\ead{carlosblanco2718@princeton.edu}

\affiliation[1]{organization={Departamento de Física},
            addressline={Pontifícia Universidade Católica do Rio de Janeiro}, 
            city={Rio de Janeiro},
            postcode={22452-970}, 
            state={RJ},
            country={Brazil}}

\affiliation[2]{organization={Department of Physics},
            addressline={Princeton University}, 
            city={Princeton},
            postcode={08544}, 
            state={NJ},
            country={USA}}
            
\affiliation[3]{organization={Stockholm University and The Oskar Klein Centre for Cosmoparticle Physics},
            addressline={Alba Nova},
            city={Stockholm},
            postcode={10691},
            country={Sweden}}



\begin{abstract}
Since the universe is not transparent to gamma rays with energies above around one hundred GeV, it is necessary to account for the interaction of high-energy photons with intergalactic radiation fields in order to model gamma-ray propagation. Here, we present a public numerical software for the modeling of gamma-ray observables. This code computes the effects on gamma-ray spectra from the development of electromagnetic cascades and cosmological redshifting. The code introduced here is based on the original $\gamma$-Cascade, and builds on it by improving its performance at high redshifts, introducing new propagation modules, and adding many more extragalactic radiation field models, which enables the ability to estimate the uncertainties inherent to EBL modeling. We compare the results of this new code to existing Monte Carlo electromagnetic transport models, finding good agreement within EBL uncertainties.
\end{abstract}




\end{frontmatter}



%

\section{Program Summary}
\noindent
Program title: $\gamma$-Cascade\\
CPC Library link to program files: \\
Developer's repository link: \url{https://github.com/GammaCascade/GCascade}\\
Licensing provisions: GPL \\
Programming language: Wolfram Language. \\
Supplementary material: Tutorial notebook, functional and auxiliary file libraries. \\
Journal Reference of previous version: JCAP 01 (2019) 013 \\
Does the new version supersede the previous version?: Yes. \\
Reasons for the new version: Improved precision, addition of several EBL models, new functions and library files. \\
Summary of revisions: $\gamma$-Cascade V4 performs more accurate electromagnetic cascade simulations, in agreement with other state-of-the-art Monte Carlo codes. Estimates of EBL modeling uncertainties on cascaded fluxes can also be made. \\
Nature of problem: Obtaining gamma-ray fluxes from astrophysical sources active above $\sim 100$~GeV is not trivial due to the absorption and regeneration of these particles en route to the Earth, producing electromagnetic cascades. Accurate simulation of cosmological cascade development typically requires the precision of numerical codes. \\
Solution method: $\gamma$-Cascade efficiently calculates cascaded fluxes from point-like and diffuse populations of sources, employing semi-analytical methods over discrete energy and redshift grids.

\section{Introduction}

Gamma rays with energies above about a hundred GeV interact significantly with cosmic photons generated by stars, dust, and the Big Bang. In specific, pair production through interactions with the cosmic microwave background (CMB) and the extragalactic background light (EBL) makes the universe non-transparent to high-energy gamma rays traveling cosmological distances. Electrons and positrons generated by these collisions can up-scatter photons from the same background radiation fields to regenerate gamma rays. This evolving cycle, known as an electromagnetic cascade, can morph the spectral distribution of gamma rays over cosmological distances \citep{Murase2012a,Berezinsky:2016feh,Venters:2010bq}.  In order to predict and understand data from gamma-ray telescopes and observatories (such as Fermi-LAT, HESS, MAGIC, VERITAS, HAWC, and CTA), a careful modeling of gamma-ray transport in the range of MeV up to PeV and above is imperative. 

$\gamma$-Cascade is an open-source software package that computes the effect of  electromagnetic cascade evolution and cosmological redshifting on the spectra of gamma rays. The work presented here significantly improves  and extends the capabilities of the original $\gamma$-Cascade by integrating a broader array of extragalactic radiation field models, incorporating methods to estimate uncertainties in the resultant spectra. This package complements existing tools like CRbeam \citep{Kalashev:2022cja} and ELMAG \citep{Blytt:2019xad} by providing a fast semi-analytical framework that is highly customizable, allowing users to modify the underlying physics model and generate new computational libraries.

The original version (V1) of $\gamma$-Cascade was introduced in 2018 and has since undergone two revisions up to V3 \citep{Blanco:2018bbf}. Here, we present a new version of $\gamma$-Cascade (V4) which is a complete overhaul of the underlying code and functionality. We have implemented four different EBL models (Saldana-Lopez \textit{et al.} (2021)~\cite{Saldana-Lopez:2020qzx}, Finke \textit{et al.} (2022)~\cite{Finke:2022uvv}, Franceschini and Rodighiero (2017)~\cite{Franceschini:2017iwq}, and Domínguez \textit{et al.} (2011)~\cite{Dominguez:2010bv}) into the code, as well as the uncertainty range for the Saldana-Lopez model in order to estimate the systematic uncertainty inherent in these calculations. We have also made significant improvements to the accuracy of our semi-analytic approach while avoiding time-consuming adaptive integration techniques. Here, we present a complete analytic treatment of electromagnetic cascades, including a general solution. We discuss the discretization of iterative solutions and their implementation in  $\gamma$-Cascade V4. This manuscript is meant to be both an exposition of the physics behind gamma-ray transport and its importance to gamma-ray observables, as well as the documentation and description of the new software package.

While other solutions exist to solve to the problem of modeling high-energy gamma-ray propagation, none provide a fast and accurate semi-analytical treatment like $\gamma$-Cascade. Existing software packages such as GALPROP~\citep{strong2009galprop,vladimirov2010galprop} and CRPropa~\citep{batista2016crpropa}  provide computational tools for the cosmological propagation of electromagnetic particles, focusing on cosmic rays. The CRbeam~\citep{Kalashev:2022cja} and ELMAG~\citep{Blytt:2019xad,kachelriess2011elmag} packages provide a cascade-centric suite of tools based on a Monte Carlo approach. Conversely, fully analytical approximations to the problem of cascade development have been proposed as in refs. \citep{Berezinsky:2016feh,Lee:1996fp,kalashev2015simulations}. In whole, all of these tools produce results in agreement with those of $\gamma$-Cascade. We present comparisons to other existing codes and analytic approximations in order to show the accuracy of  $\gamma$-Cascade V4. Finally, we highlight that in addition to point-source calculations, $\gamma$-Cascade efficiently handles the propagation of gamma rays from distributions of user-defined classes of sources in order to compute diffuse fluxes. This is possible, in principle, using other tools, but it would be significantly time-intensive since the cascades would have to be treated as a large sum of point sources. 

In this paper, we detail the fundamental processes driving gamma-ray interactions, including pair production and inverse Compton scattering. We explain the structure and functionality of the $\gamma$-Cascade libraries and main code, particularly focusing on the necessary approximations for inverse Compton scattering. We also discuss gamma-ray attenuation, the role of cascades in shaping the spectra, and the implications of cosmological propagation on gamma-ray observables. Finally, we summarize our results, compare them to other existing codes and discuss their relevance to ongoing and future gamma-ray observations.

The new $\gamma$-Cascade V4 package and its associated resources are accessible at \url{https://github.com/GammaCascade/GCascade}.

\section{Pair Production}

An electromagnetic cascade is initiated when a gamma-ray with energy $E_\gamma$ interacts with a background photon with energy $\epsilon$ (in the present case, either from the CMB or the EBL), producing an electron-positron pair (here on referred to collectively as electrons). Mass-energy conservation requires that the squared center of mass (CM) energy satisfies $s=2E_\gamma \epsilon(1-\mu) \geq (2m_e)^2 \equiv s_{\rm th}$, where $\mu\equiv \cos\theta$ is the cosine of the angle between the incoming photons. The total pair production (PP) cross section given by \citep{Breit1934bwp,Jauch1955voj}
\begin{equation}
    \sigma_{\rm PP}(s) = \sigma_{\rm T}\,\frac{3}{16}\,(1-\beta^2)\bigg[(3-\beta^4)\ln\frac{1+\beta}{1-\beta} - 2\beta(2-\beta^2)\bigg]~,
\end{equation}
where $\sigma_{\rm T}$ is the Thomson cross section and $\beta=(1-4m_e^2/s)^{1/2}$ is the velocity of the outgoing electron in the CM frame. Given an isotropic background photon field with redshift-dependent energy-differential number density $d n(\epsilon,z)/d \epsilon$, one can obtain the average interaction rate $\Gamma_{\rm PP}$ from the following,
\begin{equation}\label{eq:ppintrate}
    \Gamma_{\rm PP}(E_\gamma,z) = c\int d \epsilon \int_{-1}^{1-\frac{2m_e^2}{E_\gamma \epsilon}} d \mu \,\frac{1-\mu}{2}\,\sigma_{\rm PP}(E_\gamma,\epsilon,\mu)\,\frac{d n(\epsilon,z)}{d \epsilon}~.
\end{equation}
From Eq. (\ref{eq:ppintrate}), we note the following features of the interaction rate.
\begin{enumerate}
    \item The mean free path of gamma-rays is related to the interaction rate by $\lambda_{\rm PP} = c/\Gamma_{\rm PP}$.
    \item The upper limit on the $\mu$ integration comes from the threshold condition, $s\geq s_{\rm th}$, and is implied for all such integrals in this section.
    \item Although analytical expressions exist for the angle-averaged cross section\footnote{In this article, angle brackets ``$\langle\rangle$'' always indicate averages over the incoming particles' directions.} $\langle \sigma_{\rm PP} \rangle \equiv \int d \mu \, \sigma_{\rm PP}\,(1-\mu)/2$, they are either too cumbersome (e.g., the exact formula obtained in \citep{Gould1967ppp}) or use approximations that fail near threshold (e.g., \citep{aharonian1983}, used in $\gamma$-Cascade V3). Here, we use direct numerical integration instead.
    \item Since the CMB evolves predictably with redshift as $\frac{d n}{d \epsilon}(\epsilon,z) = (1+z)^2\, \frac{d n}{d \epsilon}(\frac{\epsilon}{1+z},0)$, the PP rate on the CMB as a function of redshift is given by,
    \begin{equation}
        \Gamma_{\rm PP}^{\rm cmb}(E_\gamma,z) = (1+z)^3 \,\Gamma_{\rm PP}^{\rm cmb}(E_\gamma(1+z),0)~.
    \end{equation}
    \item Since interaction rates are additive in the presence of multiple background photon fields, the total PP rate on the CMB and the EBL is given by, $\Gamma_{\rm PP}^{\rm tot} = \Gamma_{\rm PP}^{\rm cmb} + \Gamma_{\rm PP}^{\rm ebl}$.
\end{enumerate}

In order to obtain the spectrum of electrons $d N_e/d E'_e$ generated from the PP of an arbitrary spectrum of gamma rays $d N_\gamma/d E_\gamma$, we compute the average differential interaction rate (primed energies always refer to outgoing particles in interactions),
\begin{equation}
    \gamma_{\gamma\to e}^{\rm PP}(E_\gamma,E'_e,z) = c\int d \epsilon \int d \mu \,\frac{1-\mu}{2}\,\frac{d \sigma_{\rm PP}}{d E'_e}\,\frac{d n(\epsilon,z)}{d \epsilon}~,
\end{equation}
where we adopt the following approximate expression for the angle-averaged PP differential cross section as in previous versions of the code \citep{aharonian1983,Aharonian1985voe}:
\begin{align}
    \left\langle\frac{d \sigma_{\rm PP}}{d E'_e}\right\rangle &\equiv \int d \mu\, \frac{1-\mu}{2} \,\frac{d \sigma_{\rm PP}}{d E'_e}(E_\gamma,E'_e,\epsilon,\mu)\nonumber \\
    &= \frac{3\sigma_{\rm T}}{32}\frac{m_e^4}{E_\gamma^3 \epsilon^2}\left[\frac{4E_{\gamma}^{2}}{\left(E_{\gamma}-E'_{e}\right)E'_{e}}\ln\left(\frac{4\epsilon E'_{e}\left(E_{\gamma}-E'_{e}\right)}{m_{e}^{2}E_{\gamma}}\right) \right.\nonumber \\ & \left. \quad \, -\frac{8E_{\gamma}\epsilon}{m_{e}^{2}} + \frac{2E_{\gamma}^{2}\left(2E_{\gamma}\epsilon -m_{e}^{2}\right)}{\left(E_{\gamma}-E'_{e}\right)E'_{e}m_{e}^{2}}\right.\nonumber \\ & \left. \quad \, -\left(1-\frac{m_{e}^{2}}{E_{\gamma}\epsilon}\right)\frac{E_{\gamma}^{4}}{\left(E_{\gamma}-E'_{e}\right)^{2}E^{\prime\,2}_{e}}\right]~,
\end{align}
valid for $E_\gamma \gg m_e$. The energy of the outgoing electron is restricted to the following kinematic window,
\begin{equation}\label{eq:pp-outgoing-energy-limits}
    \frac{E_\gamma}{2}\bigg(1 - \sqrt{1-\frac{m_e^2}{E_\gamma \epsilon}}\bigg) \leq E'_e \leq \frac{E_\gamma}{2}\bigg(1 + \sqrt{1 - \frac{m_e^2}{E_\gamma \epsilon}}\bigg)~.
\end{equation}
Note that the differential interaction rate satisfies the following properties:
\begin{enumerate}
    \item \label{prop1} $\int_{E'_{e,{\rm min}}}^{E'_{e,{\rm max}}} d E'_e\, \gamma_{\gamma\to e}^{\rm PP} = \Gamma_{\rm PP}$,
    \item \label{prop2} $\int_{E'_{e,{\rm min}}}^{E'_{e,{\rm max}}} d E'_e\, E'_e\,\gamma_{\gamma\to e}^{\rm PP}/\Gamma_{\rm PP} = \langle E'_e\rangle = E_\gamma/2$,
    \item $\gamma_{\gamma\to e}^{\rm PP,cmb}(E_\gamma,E'_e,z) = (1+z)^4\, \gamma_{\gamma\to e}^{\rm PP,cmb}(E_\gamma(1+z),E'_e(1+z),0)$, for the CMB, and
    \item $\gamma_{\gamma\to e}^{\rm PP,tot} = \gamma_{\gamma\to e}^{\rm PP,cmb}+\gamma_{\gamma\to e}^{\rm PP,ebl}$, just as in the additive property of $\Gamma_{\rm PP}$.
\end{enumerate}
We suppress the integration limits of all future integrals over the outgoing particles' energies, which are implied from the allowed kinematic ranges such as the one in Eq. (\ref{eq:pp-outgoing-energy-limits}). 

From these interaction rates, we obtain the properly normalized spectrum of pair-produced electrons from monoenergetic gamma rays of energy $E_\gamma$,
\begin{equation}\label{eq:ppspec-monoenergetic}
    \frac{d N_{\gamma\to e}}{d E'_e}(E'_e,E_\gamma,z) = 2 \,\frac{\gamma_{\gamma\to e}^{\rm PP}(E_\gamma,E'_e,z)}{\Gamma_{\rm PP}(E_\gamma,z)}~.
\end{equation}
The factor of 2 accounts for both electrons and positrons being produced with symmetric probability distributions. Property \ref{prop1} above ensures particle number is fixed,
\begin{equation}\label{eq:pp-number-cons}
    \int dE'_e \,\frac{d N_{\gamma\to e}}{d E'_e}(E'_e,E_\gamma,z) = 2~,
\end{equation}
while property \ref{prop2} guarantees energy conservation,
\begin{equation}\label{eq:pp-energy-cons}
    \int dE'_e \,E'_e\,\frac{d N_{\gamma\to e}}{d E'_e}(E'_e,E_\gamma,z) = E_\gamma~.
\end{equation}
It can be useful to think of Eq. (\ref{eq:pp-number-cons}) as a normalization condition for the spectrum, while Eq. (\ref{eq:pp-energy-cons}) can be seen as the expectation value for the energy of outgoing electrons, 
\begin{equation}
    \langle E'_e\rangle = \frac{\int dE'_e \,E'_e\,\frac{d N_{\gamma\to e}}{d E'_e}}{\int dE'_e \,\frac{d N_{\gamma\to e}}{d E'_e}} = \frac{E_\gamma}{2}~.
\end{equation}

These normalized spectra are akin to a Green's function in energy space that solves the cascade problem. Therefore, for an arbitrary spectrum of gamma rays, $d N_\gamma/d E_\gamma$, the resulting $e^\pm$ spectrum is given by,
\begin{equation}\label{eq:pp-arbitrary-spec}
    \frac{d N_e}{d E'_e}(E'_e,z) = \int d E_\gamma \, \frac{d N_{\gamma\to e}}{d E'_e}(E'_e,E_\gamma,z)\,\frac{d N_\gamma}{d E_\gamma}(E_\gamma,z)~.
\end{equation}
Props. \ref{prop1} and \ref{prop2} guarantee energy and particle number conservation for any such spectrum, which is an essential requirement for proper development of electromagnetic cascades.

\section{Inverse Compton Scattering}

After PP, the resulting electrons of energy $E_e$ can undergo inverse Compton scattering (ICS) off of target background photons, primarily from the CMB, generating new high energy gamma rays. The cross section for this process is given by
\begin{multline}\label{eq:sigmaics}
    \sigma_{\rm ICS}(s) = \frac{3\sigma_{\rm T}}{8}\frac{m_e^2}{s\beta}\left[\frac{2}{\beta(1+\beta)}(2+2\beta-\beta^2-2\beta^3)\right. \\
    \left. - \frac{1}{\beta^2}(2-3\beta^2-\beta^3)\ln\frac{s}{m_e^2}\right]~,
\end{multline}
where $s=m_e^2+2E_e \epsilon(1-\beta_e \mu)$ is the squared CM energy, in terms of the incoming electron's velocity $\beta_e = (1-m_e^2/E_e^2)^{1/2} \approx 1$ and the cosine of the angle between the incoming particles $\mu$, and $\beta = (s-m_e^2)/(s+m_e^2)$ is the velocity of the outgoing electron in the CM frame. With Eq. (\ref{eq:sigmaics}), we can calculate interaction rates $\Gamma_{\rm ICS}$ (or mean free paths $\lambda_{\rm ICS}$) in a similar fashion to Eq. (\ref{eq:ppintrate}), with only two subtle changes required when taking angular ($\mu$) averages: \textit{(i)} the flux factor becomes $(1-\mu)/2 \to (1-\beta_e\mu)/2$, which leads to negligible corrections, and \textit{(ii)} the upper limit of $\mu$ integrations is now 1, since there is no kinematic threshold. Once again, we perform a direct numerical integration for $\langle\sigma_{\rm ICS}\rangle$, rather than using approximate analytical expressions (e.g., \citep{Zdziarski1988oij}).

Unlike in PP, where the outgoing particles have symmetric distributions, we have two distinct differential interaction rates for ICS with respect to the energy of each outgoing particle $E'_\gamma$ and $E'_e$:
\begin{equation}
    \gamma_{e\to \gamma}^{\rm ICS}(E_e,E'_\gamma,z) = c\int d \epsilon \int d \mu \,\frac{1-\beta_e\mu}{2}\,\frac{d \sigma_{\rm ICS}}{d E'_\gamma}\,\frac{d n(\epsilon,z)}{d \epsilon}~,
\end{equation}
\begin{equation}
    \gamma_{e\to e}^{\rm ICS}(E_e,E'_e,z) = c\int d \epsilon \int d \mu \,\frac{1-\beta_e\mu}{2}\,\frac{d \sigma_{\rm ICS}}{d E'_e}\,\frac{d n(\epsilon,z)}{d \epsilon}~.
\end{equation}
The first differential cross section (valid in the limit $E'_\gamma \gg \epsilon$ and $E_e \gg m_e$) is given by \citep{aharonian1981,Blumenthal:1970gc}
\begin{align}\label{eq:icsdiffxsec}
    \left\langle\frac{d \sigma_{\rm ICS}}{d E'_\gamma}\right\rangle &\equiv \int d \mu\, \frac{1-\beta_e \mu}{2} \,\frac{d \sigma_{\rm ICS}}{d E'_\gamma}(E_e,E'_e,\epsilon,\mu)\nonumber \\
    &= \frac{3\sigma_{\rm T}m_{e}^{2}}{4\epsilon E_{e}^{2}}\left[1+\frac{z^{2}}{2\left(1-z\right)}+\frac{z}{b\left(1-z\right)}-\frac{2z^{2}}{b^{2}\left(1-z\right)^2}\right.\nonumber\\
    &\quad \, -  \frac{z^{3}}{2b\left(1-z\right)^{2}}-\frac{2z}{b\left(1-z\right)}\,\ln\frac{b\left(1-z\right)}{z} \bigg]~,
\end{align}
where $b\equiv 4\epsilon E_e/m_e^2$, $z \equiv E'_\gamma/E_e$, and the kinematically allowed range is constrained to $\epsilon/(E_e+\epsilon) \leq z\leq b/(1+b)$. By substituting $E'_\gamma \to E_e - E'_e$ in Eq. (\ref{eq:icsdiffxsec}), we obtain the differential cross section with respect to the outgoing electron energy $\langle d\sigma_{\rm ICS}/dE'_e \rangle$. We compute the normalized spectra of inverse-Compton-scattered photons and electrons from incoming electrons with a fixed energy $E_e$ as
\begin{equation}\label{eq:ics-spec-monoenergetic}
    \frac{dN_{e\to \gamma}}{dE'_\gamma}(E'_\gamma,E_e,z) = \frac{\gamma_{e\to \gamma}^{\rm ICS}(E_e,E'_\gamma,z)}{\Gamma_{\rm ICS}(E_e,z)}~,
\end{equation}
\begin{equation}
    \frac{dN_{e\to e}}{dE'_e}(E'_e,E_e,z) = \frac{\gamma_{e\to e}^{\rm ICS}(E_e,E'_e,z)}{\Gamma_{\rm ICS}(E_e,z)}~,
\end{equation}
which satisfy analogous particle number and energy conservation conditions to Eqs. (\ref{eq:pp-number-cons}) and (\ref{eq:pp-energy-cons}),
\begin{equation}\label{eq:ics-number-cons}
    \int dE'_\gamma \,\frac{d N_{e\to \gamma}}{d E'_\gamma} = \int dE'_e \,\frac{d N_{e\to e}}{d E'_e} = 1~,
\end{equation}
\begin{equation}\label{eq:ics-energy-cons}
    \int dE'_\gamma \,E'_\gamma\,\frac{d N_{e\to \gamma}}{d E'_\gamma} + \int dE'_e \,E'_e\,\frac{d N_{e\to e}}{d E'_e} = E_e~,
\end{equation}
 respectively. We can obtain outgoing gamma-ray and electron spectra from an arbitrary incoming spectrum of electrons by integrating over the differential electron spectrum just as in Eq. (\ref{eq:pp-arbitrary-spec}).

\section{The $\gamma$-Cascade Library}

The $\gamma$-Cascade code is written entirely in Wolfram language, intended for use in Mathematica. All of its variables and functions are contained inside one package file: \texttt{GCascadeV4.wl}. However, $\gamma$-Cascade also contains a library with several pre-computed tables necessary for the code's functionality which contain the numerical results of some key integrals above, e.g. the results of normalized PP and ICS spectra. These tables are evaluated at discrete energy and redshift values, listed in the arrays ``\texttt{energies}'' and ``\texttt{zReg}'' respectively. The former contains a list of 300 energies between $10^{-1}$--$10^{12}$~GeV, evenly spaced in a logarithmic scale, and is available to the user in order to generate injection spectra that is readable to the code. Meanwhile, the latter is internal to the code and contains a list of 1001 redshifts from 0 to 10, equally spaced in intervals of 0.01. 

Since the files containing these tables can be quite large, we have divided the library into two parts for the user's convenience. Those which are essential for code's functionality come together with the main $\gamma$-Cascade V4 package and include:

\begin{itemize}
    \item PP inverse mean free paths (IMFPs), $\lambda^{-1}_{\rm PP} = \Gamma_{\rm PP}/c$, in units of [cm$^{-1}$], for interactions with the CMB and each of the EBL models available in the code. These tables have dimensions $1001 \times 300$ corresponding to the redshifts and energies in the \texttt{zReg} and \texttt{energies} arrays, respectively. In other words, the entry $ij$ of each IMFP array contains $\lambda^{-1}_{\rm PP}(E_j,z_i)$. They can be found inside the ``pp-IMFPs'' directory.
    
    \item Normalized electron spectra from PP of monoenergetic gamma rays, $dN_{\gamma\to e}/dE'_e$, in units of [eV$^{-1}$], from interactions with different background photon fields. These tables have dimensions $1001 \times 300 \times 300$ corresponding to the entries in \texttt{zReg} and \texttt{energies} (twice). Two energy dimensions are necessary for the incoming particle (unprimed) and the outgoing particle (primed): $dN_{\gamma\to e}/dE'_e (E'_k,E_j,z_i)$ is the entry $ijk$ from each array. After evaluating Eq. (\ref{eq:ppspec-monoenergetic}) for each combination of energies and redshift, the tables were also post-processed to ensure energy conservation under trapezoidal integration on the discrete grid, as described in Appendix \ref{app:grid}. They can be found inside the ``pp-spec'' directory and are only necessary for changing the default magnetic field in $\gamma$-Cascade.

    \item On-the-spot ICS spectra from monoenergetic electrons, $dN_{e\to \gamma,{\rm OTS}}/dE'_\gamma$, in [eV$^{-1}$], calculated using numerical techniques described in Section \ref{sec:on-the-spot} and Appendix \ref{app:on-the-spot}. These are also necessary only for changing the magnetic field. Array dimensions are $1001 \times 300 \times 300$ as before, and are found in the ``on-the-spot-ics-spec'' directory.
 
    \item Normalized gamma-ray spectra after an on-the-spot PP-ICS cycle initiated by monochromatic gamma rays, $dN_{\gamma \to e \to \gamma,{\rm OTS}}/dE'_\gamma$, defined by Eq. (\ref{eq:ots-monochrom-spec}) and in units of [eV$^{-1}$]. These are the core cumulative result of our numerical treatment and are imported by the $\gamma$-Cascade code to evaluate cascaded spectra using Eqs. (\ref{eq:pp-ics-cycle-spec}) and (\ref{eq:generic-ots-spec}). These arrays are also $1001 \times 300 \times 300$ and can be found inside the ``cycle-spec'' directory. They include effects of magnetic fields described at the end of Section \ref{sec:on-the-spot}.

    \item ICS energy loss rates, $(dE/dt)_{\rm ICS}$, in [eV$\,$s$^{-1}$], obtained by evaluating Eq. (\ref{eq:ics-E-loss-rate}) over the same $1001 \times 300$ grid as the PP IMFPs. These are also imported by the code whenever the user changes the magnetic field, in order to calculate $f_{\rm ICS}$ via Eq. (\ref{eq:fICS}). They are available in the directory ``ics-E-loss-rates''.
  
    \item Auxiliary arrays \texttt{stepSizeArray.mat} and \texttt{zRegIndexArray.mat}. The former contains light-travel distances in [Mpc] in redshift steps of $10^{-6}$ arranged in a particular structure for internal use in $\gamma$-Cascade. The latter contains indexing conversions between different redshift tables inside the code.

    \item A blank text file \texttt{BfieldChangelog.txt} used for keeping track of changes made to $\gamma$-Cascade's default magnetic field.
\end{itemize}
Non-essential tables can be found in an auxiliary library space, also available at \url{https://github.com/GammaCascade/GCascade}. These include:
\begin{itemize}
    \item ICS IMFP tables, structured just like the previously described PP IMFP tables, and found inside the ``ics-IMFPs'' directory.
  
    \item Differential IMFPs, $\gamma^{\rm PP/ICS}(E_j,E'_k,z_i)/c$, in units of [cm$^{-1}$ eV$^{-1}$], for PP and ICS with the CMB or the EBL, over the same $1001 \times 300 \times 300$ grid as before. They can be found inside the ``pp-differential-IMFPs'' and ``ics-differential-IMFPs'' directories. Note that ICS differential IMFPs can be with respect to either the outgoing photons or electrons, and are labeled accordingly with ``gamma'' or ``e'' at the end. 

    \item Normalized gamma-ray and electron spectra from ICS of monoenergetic electrons, in [eV$^{-1}$]. These contain post-processed $dN_{e \to \gamma/e}/dE'_{\gamma/e}(E'_k,E_j,z_i)$ values, similar to the PP spectral tables, as described in Appendix \ref{app:grid}. They can be found in the ``ics-spec'' directory.

    \item Modified ICS spectral grids, $dN_{e \to \gamma/e}/dE'_{\gamma/e}$, after further processing as described in Appendix \ref{app:on-the-spot}. These were used for calculating on-the-spot spectra, and can be found in the directory ``modified-ics-spec''.

    \item Gamma-ray spectra from PP-ICS cycles, $dN_{\gamma \to e \to \gamma,{\rm OTS}}/dE'_\gamma$, similar to before, but without including intergalactic magnetic field effects (IGMF; i.e. assuming $B=0$). These are available in the ``cycle-spec-B0'' directory.
\end{itemize}

The fact that all of the physical quantities and operations within $\gamma$-Cascade are evaluated on a grid invites an intuitive tensor-like interpretation. Interaction rates/IMFPs on a grid are denoted by $\Gamma_{ij} \equiv \Gamma(E_j,z_i) = \lambda^{-1}(E_j,z_i) \equiv \lambda^{-1}_{ij}$, differential interaction rates are denoted by $\gamma_{ijk} \equiv \gamma(E_j,E'_k,z_i)$, and so on. The order of the indices follows the order they appear in the array dimensions, with $i$ always representing the $i$'th redshift entry in the \texttt{zReg} array. This allows us to recast equations from the previous two sections into index notation. For example, Eq. (\ref{eq:ppspec-monoenergetic}) on a grid looks like
\begin{equation}
    \left(\frac{dN_{\gamma\to e}}{dE'_e}\right)_{ijk} = 2\frac{(\gamma^{\rm PP}_{\gamma\to e})_{ijk}}{(\Gamma_{\rm PP})_{ij}}~,
\end{equation}
and integrals such as Eq. (\ref{eq:pp-number-cons}) can be done by trapezoidal integration, which reduces to computationally-efficient matrix multiplication and addition,
\begin{equation}\label{eq:pp-number-cons-trapint}
    \sum_{k=1}^{299} \left[\left(\frac{dN_{\gamma\to e}}{dE'_e}\right)_{i,j,k+1} + \left(\frac{dN_{\gamma\to e}}{dE'_e}\right)_{i,j,k} \right]\frac{E_{k+1} - E_k}{2} = 2~, \quad \forall~i,j~.
\end{equation}
In particular, the discrete version of spectral integrals such as Eq. (\ref{eq:pp-arbitrary-spec}) is evaluated in $\gamma$-Cascade by trapezoidal integration rather than adaptive sampling of the continuous integral, thereby saving a significant amount of computational time:
\begin{multline}\label{eq:pp-arbitrary-spec-trapint}
    \left(\frac{dN_e}{dE'_e}\right)_{i,k} = \sum_{j=1}^{299} \left[\left(\frac{dN_{\gamma\to e}}{dE'_e}\right)_{i,j+1,k} \left(\frac{dN_\gamma}{dE_\gamma}\right)_{i,j+1} \right.\\+ \left. \left(\frac{dN_{\gamma\to e}}{dE'_e}\right)_{i,j,k}\left(\frac{dN_\gamma}{dE_\gamma}\right)_{i,j} \right] \frac{E_{j+1} - E_j}{2}~.
\end{multline}
This proves to be extremely useful when evaluating spectra after multiple PP/ICS steps, which would normally require nested numerical integration in the continuous realm.

\section{Structure of the Main Code}

The main code \texttt{GCascadeV4.wl} contains several functions that perform gamma-ray propagation, modify magnetic fields and switch between EBL models. They are described in Appendix \ref{app:functions}, as well as in the \texttt{Tutorial.nb} notebook that comes with the $\gamma$-Cascade package. All of the propagation modules have a similar structure.

Distances in $\gamma$-Cascade are divided into redshift steps of $\Delta z$ as defined in the array called \texttt{diffuseDistances}, which contains redshift values from $10^{-6}$ -- $10^1$ in non-uniform spacings, varying between $10^{-6} \leq \Delta z \leq 10^{-2}$. Each of these intervals is further subdivided into fine redshift steps of $\delta z = 10^{-6}$. During the propagation of gamma rays, spectra may get attenuated by PP and regenerated by ICS at every $\delta z$ step, which we call a ``cycle''.

For an incoming spectrum $dN_\gamma/dE_\gamma$, the attenuated spectrum after one cycle is given by
\begin{equation}
    \left(\frac{dN_\gamma}{dE_\gamma} (E_\gamma)\right)_{\rm att} = e^{-\tau_{\rm PP}(E_\gamma)}\frac{dN_{\rm \gamma}}{dE_\gamma}(E_\gamma)~,
\end{equation}
where the optical depth along a light-travel length $\ell = \int_{z_i}^{z_{i+1}} dz\,[c/(1+z)H(z)]$ corresponding to the redshift interval $\delta z = z_{i+1} - z_i$ is given by the following,
\begin{align}
    \tau_{\rm PP}(E_\gamma) &= \int_{z_i}^{z_{i+1}} dz\,\frac{c}{(1+z)H(z)} \int_{m_e^2/E_\gamma}^\infty d\epsilon\,\frac{dn(\epsilon,z)}{d\epsilon}\,\langle \sigma_{\rm PP}\rangle (E_\gamma,\epsilon) \nonumber\\
    &\approx \frac{\ell}{\lambda_{\rm PP}(E_\gamma,z^*)}~,
\end{align}
where $H(z)=H_0\sqrt{\Omega_m (1+z)^3 + \Omega_\Lambda}$ is the Hubble parameter, with $H_0 = 67.4$~km\,s$^{-1}$\,Mpc$^{-1}$, $\Omega_m = 0.315$, and $\Omega_\Lambda = 0.685$ \citep{Planck:2018vyg}, and $\lambda_{\rm PP}$ is the mean free path as a function of the energy $E_\gamma$ in the lab (not comoving) frame. Since $\lambda_{\rm PP}$ is evaluated on a grid of redshifts determined by the \texttt{zReg} array, $\gamma$-Cascade uses its value at the closest entry to $z_{i}$, represented by $z^*$, as indexed by the \texttt{zRegIndexArray.mat} array. The remaining part of the spectrum, $(1-e^{-\tau_{\rm PP}}) dN_\gamma/dE_\gamma$ undergoes PP and may be added back into the spectrum after ICS. In $\gamma$-Cascade, electrons are assumed to lose their entire energy through successive ICS interactions within a single cycle. We call this the \textit{on-the-spot approximation}, which we describe in detail on the next section.

These cycles repeat until they reach the end of a $\Delta z = z_2 - z_1$ window, where cosmological redshifting is applied to the energies in the spectrum:
\begin{equation}
    E \to E\times \frac{1 + z_1}{1+z_2}~.
\end{equation}
The process repeats for all subsequent $\Delta z$ windows until the cascade terminates at $z=0$. For point sources, a normalization factor of $(1+z_{\rm source})^2$ is necessary in order to reflect the redshifting of the (energy$^{-1}$ time$^{-1}$) units of $dN/dE$. The spectrum is multiplied by this factor at the end of the cascade calculation. In order to turn the luminosity into a flux, the spectrum is divided by the light-sphere surface area $(4\pi d_L^2)$, where $d_L$ is the luminosity distance to the source. In the case of a distribution of sources, a new injected contribution is added at each new $\Delta z$ and normalized through the comoving volume element $\Delta \mathcal{V}_c (z)$.

\subsection{On-The-Spot Approximation for Inverse Compton Scattering}
\label{sec:on-the-spot}

\begin{figure*}[t!]
    \centering
    \includegraphics[width=\textwidth]{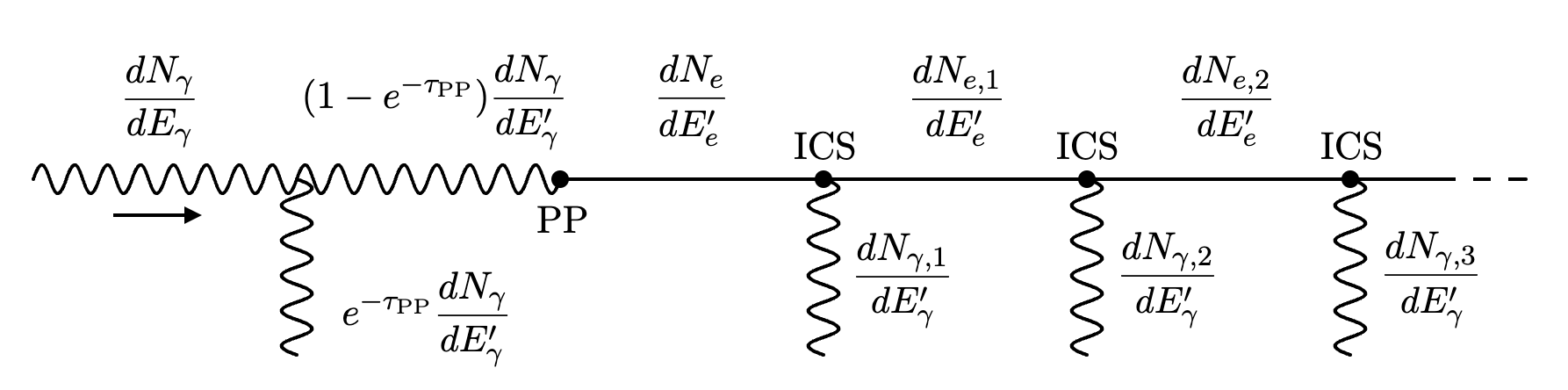}
    \caption{Schematic representation of an interaction cycle in $\gamma$-Cascade. Besides the gamma rays that do not undergo PP, $e^{-\tau_{\rm PP}}(dN_\gamma/dE'_\gamma)$, the next cycle receives photons from successive ICSs of electron spectra, $\sum_n dN_{\gamma,n}/dE'_\gamma$, according to our on-the-spot approximation.}
    \label{fig:on-the-spot-scheme}
\end{figure*}

$\gamma$-Cascade adopts an on-the-spot approximation in which all electrons produced at a given redshift undergo infinitely many ICS interactions, losing all their energy in the form of photons, at that same redshift. As a consequence, the gamma-ray spectrum exiting each interaction cycle is the sum of the attenuated spectrum (i.e. the surviving photons which did not undergo PP in that cycle), the gamma rays from the ICS of the pair-produced electrons, the gamma rays from the ICS of the electrons produced after the first ICS, and so on. This is represented schematically in Figure \ref{fig:on-the-spot-scheme}, showing what happens inside each interaction cycle within $\gamma$-Cascade. Following its notation, the gamma-ray spectrum leaving the interaction cycle after PP and infinitely many ICSs is\footnote{$E'_\gamma$ represents the energy of gamma rays leaving the interaction cycle, while $E_\gamma$ refers to those entering it.}
\begin{equation}\label{eq:pp-ics-cycle-spec}
    \frac{dN_{\gamma,tot}}{dE'_\gamma}(E'_\gamma,z) = \bigg[e^{-\tau_{\rm PP}(E_\gamma,\delta z)}\frac{dN_\gamma}{dE_\gamma}(E_\gamma,z)\bigg]\bigg|_{E_\gamma = E'_\gamma} + \frac{dN_{\gamma,{\rm OTS}}}{dE'_\gamma}(E'_\gamma,z)~,
\end{equation}
where the first term represents the surviving gamma rays that do not undergo PP, and the second term represents the sum of all on-the-spot ICS contributions, given by
\begin{equation}\label{eq:OTS-cycle-spec}
    \frac{dN_{\gamma,{\rm OTS}}}{dE'_\gamma}(E'_\gamma,z) = \sum_{n=1}^\infty \frac{dN_{\gamma,n}}{dE'_\gamma}(E'_\gamma,z)~.
\end{equation}

The electron spectrum $dN_e/dE'_e$ arising from the partial PP of a generic incoming gamma-ray flux $dN_\gamma/dE_\gamma$, as in Figure \ref{fig:on-the-spot-scheme}, is obtained analogously to Eq. (\ref{eq:pp-arbitrary-spec}):
\begin{multline}\label{eq:partial-elec-spec}
    \frac{dN_e}{dE'_e}(E'_e,z) = \int dE_\gamma \,\Bigg\{ \frac{dN_{\gamma \to e}}{dE'_e}(E'_e,E_\gamma,z)\,\\ \times \bigg[\Big(1 - e^{-\tau_{\rm PP}(E_\gamma)}\Big)\frac{dN_\gamma}{dE_\gamma}(E_\gamma,z)\bigg] \Bigg\}~.
\end{multline}
We calculate $dN_{\gamma,{\rm OTS}}/dE'_\gamma$, given a generic electron spectrum from Eq. (\ref{eq:partial-elec-spec}), by first considering the normalized photon spectra from on-the-spot ICSs of monoenergetic electrons, $dN_{e\to \gamma,{\rm OTS}}/dE'_\gamma$. The resulting gamma-ray spectrum is given by the following,
\begin{equation}\label{eq:ots-spec-from-generic-elecspec}
    \frac{dN_{\gamma,{\rm OTS}}}{dE'_\gamma}(E'_\gamma,z) = \int dE'_e \,\frac{dN_{e\to \gamma,{\rm OTS}}}{dE'_\gamma}(E'_\gamma,E'_e,z)\,\frac{dN_e}{dE'_e}(E'_e,z)~.
\end{equation}
We can relate $dN_{\gamma,{\rm OTS}}/dE'_\gamma$ directly to the incoming gamma-ray spectrum through the use of Eq. (\ref{eq:partial-elec-spec}),
\begin{multline}\label{eq:generic-ots-spec}
    \frac{dN_{\gamma,{\rm OTS}}}{dE'_\gamma}(E'_\gamma,z) = \int dE_\gamma \, \Bigg\{\frac{dN_{\gamma\to e \to \gamma,{\rm OTS}}}{dE'_\gamma}(E'_\gamma,E_\gamma,z)\,\\ \times \bigg[\Big(1 - e^{-\tau_{\rm PP}(E_\gamma)}\Big)\frac{dN_\gamma}{dE_\gamma}(E_\gamma,z)\bigg]\Bigg\}~,
\end{multline}
where
\begin{multline}\label{eq:ots-monochrom-spec}
    \frac{dN_{\gamma\to e \to \gamma,{\rm OTS}}}{dE'_\gamma}(E'_\gamma,E_\gamma,z) = \\ \int dE'_e \,\frac{dN_{\gamma \to e}}{dE'_e}(E'_e,E_\gamma,z)\,\frac{dN_{e\to \gamma,{\rm OTS}}}{dE'_\gamma}(E'_\gamma,E'_e,z)~.
\end{multline}

We now turn to finding $dN_{e\to \gamma,{\rm OTS}}/dE'_\gamma$. As in Figure \ref{fig:on-the-spot-scheme}, it is given by the contribution of all on-the-spot ICS generations $dN_{e\to\gamma,n}/dE'_\gamma$, this time coming from an initial monoenergetic electron of energy $E_e$ (rather than from a generic spectrum as in $dN_{\gamma,n}/dE'_\gamma$ in Figure \ref{fig:on-the-spot-scheme}),
\begin{equation}\label{eq:on-the-spot-spec}
    \frac{dN_{e\to \gamma,{\rm OTS}}}{dE'_\gamma}(E'_\gamma,E_e,z) = \sum_{n=1}^\infty \frac{dN_{e\to \gamma,n}}{dE'_\gamma}(E'_\gamma,E_e,z)~.
\end{equation}
This is essentially the Green's function version of Eq. (\ref{eq:OTS-cycle-spec}). Note that we have dropped the prime in $E'_e$ from Eq. (\ref{eq:ots-monochrom-spec}), not only for convenience, but also because we now shift our perspective to consider the electrons as the incoming particles. The first generations of photons and electrons after ICS are given by Eqs. (\ref{eq:ics-spec-monoenergetic}),
\begin{equation}\label{eq:on-the-spot-gen1}
    \frac{dN_{e\to \gamma,1}}{dE'_\gamma}(E'_\gamma,E_e,z) = \frac{dN_{e\to \gamma}}{dE'_\gamma}(E'_\gamma,E_e,z)~,
\end{equation}
\begin{equation}
    \frac{dN_{e\to e,1}}{dE'_e}(E'_e,E_e,z) = \frac{dN_{e\to e}}{dE'_e}(E'_e,E_e,z)~,
\end{equation}
and the subsequent generations have spectra given by
\begin{multline}\label{eq:on-the-spot-nth-gammaspec}
    \frac{dN_{e\to \gamma,n}}{dE'_\gamma}(E'_\gamma,E_e,z) =\\ \int dE''_e \,\frac{dN_{e\to \gamma}}{dE'_\gamma}(E'_\gamma,E''_e,z)\,\frac{dN_{e\to e,n-1}}{dE'_e}(E''_e,E_e,z)~,
\end{multline}
\begin{multline}\label{eq:on-the-spot-nth-espec}
    \frac{dN_{e\to e,n}}{dE'_e}(E'_e,E_e,z) =\\ \int dE''_e \,\frac{dN_{e\to e}}{dE'_e}(E'_e,E''_e,z)\,\frac{dN_{e\to e,n-1}}{dE''_e}(E''_e,E_e,z)~.
\end{multline}
By induction, it can be shown that the electron particle number stays constant,
\begin{equation}
    \int dE'_e \,\frac{dN_{e\to e,n}}{dE'_e}(E'_e,E_e,z) = 1~,
\end{equation}
while the total photon particle number grows with each generation, 
\begin{multline}
    \int dE'_\gamma \,\frac{dN_{e\to \gamma,n}}{dE'_\gamma}(E'_\gamma,E_e,z) = 1 \\ \Rightarrow \qquad \int dE'_\gamma \,\frac{dN_{e\to \gamma,{\rm OTS}}}{dE'_\gamma}(E'_\gamma,E_e,z) = N~,
\end{multline}
where $N$ is the number of generations included in the summation (\ref{eq:on-the-spot-spec}), tending to infinity. Energy conservation can also be guaranteed as follows,
\begin{multline}
    \int dE'_\gamma \,E'_\gamma \,\frac{dN_{e\to \gamma,n}}{dE'_\gamma}(E'_\gamma,E_e,z) + \int dE'_e \,E'_e \,\frac{dN_{e\to e,n}}{dE'_e}(E'_e,E_e,z)  \\ = \int dE''_e \,E''_e \,\frac{dN_{e\to e,n-1}}{dE''_e}(E''_e,E_e,z)~.
\end{multline}

There is an elegant way of writing the solution to this problem. Eq. (\ref{eq:on-the-spot-spec}) can be rewritten by using expressions (\ref{eq:on-the-spot-nth-gammaspec}) and (\ref{eq:on-the-spot-nth-espec}), to give an integral equation for the total ICS gamma rays,
\begin{equation}
    \frac{dN_{e\to \gamma,{\rm OTS}}}{dE'_\gamma}(E'_\gamma,E_e,z) = \int dE'_e \,\frac{dN_{e\to\gamma}}{dE'_\gamma}(E'_\gamma,E'_e,z)\,F(E'_e,E_e,z)~,
\end{equation}
where the function $F$ satisfies the Fredholm equation of the second kind
\begin{multline}\label{eq:fredholm}
    F(E'_e,E_e,z) = \delta(E_e-E'_e) \\+ \int dE''_e \,\frac{dN_{e\to e}}{dE'_e}(E'_e,E''_e,z)\,F(E''_e,E_e,z)~,
\end{multline}
with $dN_{e\to e}/dE'_e$ as the kernel function. Solving Eq. (\ref{eq:fredholm}) for $F$ would require the need of performing nested cascade integrations increasing in depth with each generation. Unfortunately, Eq. (\ref{eq:fredholm}) can only be solved recursively, so there is no direct way of obtaining an analytical expression for $F$, and thus for $dN_{e\to \gamma,{\rm OTS}}/dE'_\gamma$.

We precompute the numerical integrations on the $\gamma$-Cascade grid via trapezoidal integration. In particular, Eqs. (\ref{eq:on-the-spot-nth-gammaspec}) and (\ref{eq:on-the-spot-nth-espec}) can be rewritten in a similar fashion to Eq. (\ref{eq:pp-arbitrary-spec-trapint}) for fast evaluation, such that we can solve the problem iteratively up to whatever generation is needed to converge the resulting on-the-spot spectra, i.e. cascade completion. In general, the changing nature of the spectra calls for adaptive techniques to maintain accuracy. However, since we know how these shapes should change, we adopt a method for post-processing the spectra (described in detail in Appendix \ref{app:on-the-spot}) in order to ensure accuracy without the need for time-intensive adaptive integration schemes. Using Eq. (\ref{eq:ots-monochrom-spec}), we calculate the $dN_{\gamma \to e \to \gamma,{\rm OTS}}/dE'_\gamma$ grid from $dN_{e\to \gamma,{\rm OTS}}/dE'_\gamma$.  This last grid is the one used in $\gamma$-Cascade to calculate the photon output, Eq. (\ref{eq:pp-ics-cycle-spec}), in a given redshift step $\delta z$, after the cycle illustrated in Figure \ref{fig:on-the-spot-scheme}.

\begin{figure}[t!]
    \centering
    \includegraphics[width=0.48\textwidth]{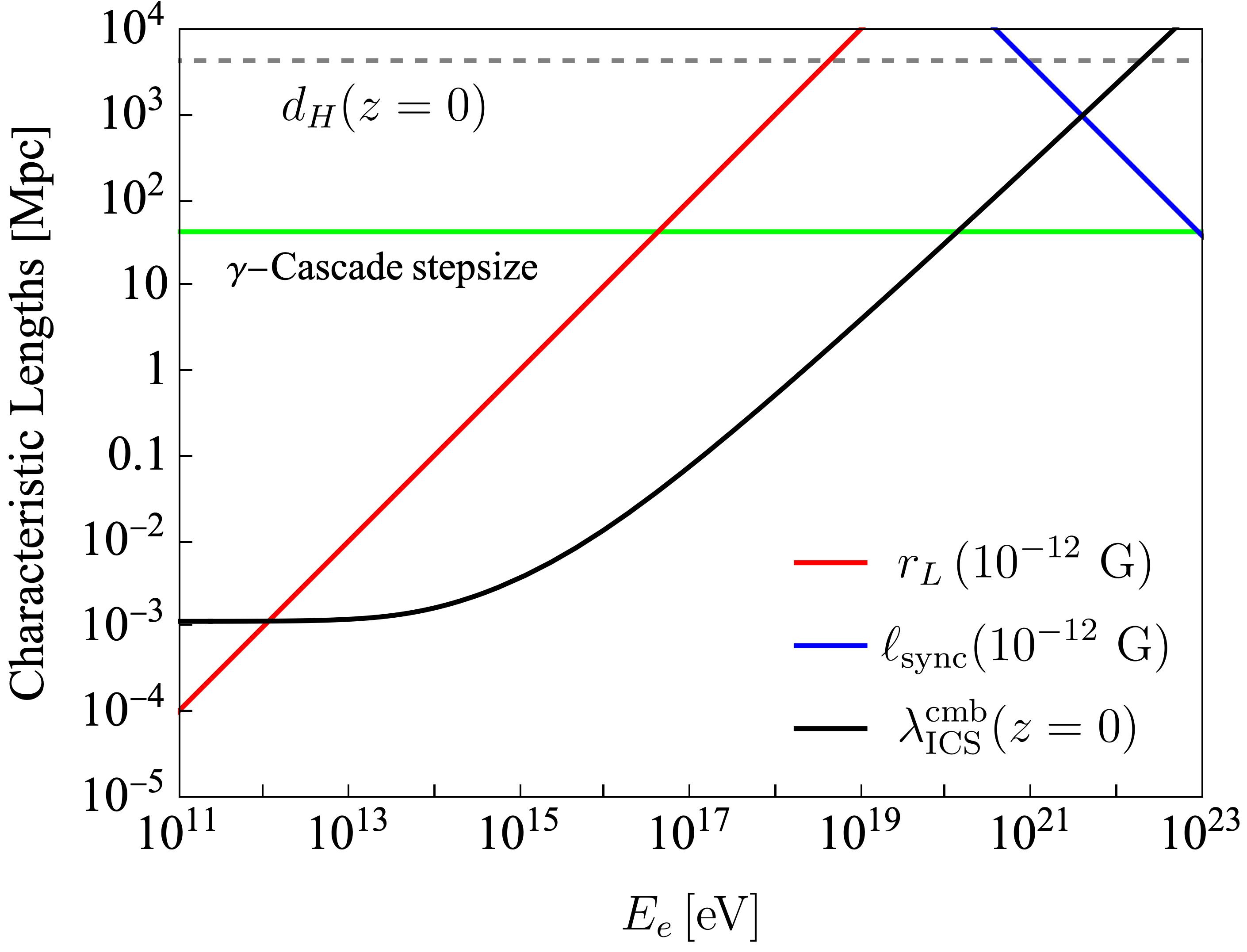}
    \caption{Characteristic lengths for electrons in the cascade. The black curve shows their mean free paths for ICS at redshift zero. For an IGMF of $B=10^{-12}$, we also show their Larmor radii in red and their synchrotron energy loss lengths in blue. The green line represents the distance scale in $\gamma$-Cascade where photon backgrounds are assumed to be constant. The gray dashed line is the Hubble length at $z=0$.}
    \label{fig:gcasc-lengths}
\end{figure}

We note the following points regarding the limitations of $\gamma$-Cascade and the validity of the on-the-spot approximation. Although the on-the-spot condition seems like a drastic assumption, it yields accurate results in practically all scenarios. This can be understood by comparing the characteristic lengths of the relevant processes in electromagnetic cascades, shown in Figure \ref{fig:gcasc-lengths}. We show the ICS mean free path at $z=0$, the Larmor radius $r_L = p/eB$ and the synchrotron energy-loss length $\ell_{\rm sync} = cE/|dE/dt|_{\rm sync}$, where $|dE/dt|_{\rm sync} = 4\sigma_{\rm T}u_B p^2/3m_e^2c$ is the energy-loss rate and $u_B = B^2/2\mu_0$ is the magnetic field energy density (in SI units), for an IGMF of $B = 10^{-12}$~G. The important length scale intrinsic to $\gamma$-Cascade is marked as the green line and labeled as ``$\gamma$-Cascade stepsize''. Although the individual PP-ICS cycles occur in redshift intervals of $\delta z = 10^{-6}$, $\gamma$-Cascade only applies redshift updates to the CMB/EBL backgrounds in steps of $\Delta z = 0.01$, corresponding to the entries on the \texttt{zReg} array. The green line in Figure \ref{fig:gcasc-lengths} specifically marks the light-travel distance between $0<z<0.01$, which all particles cross. The following conclusions can be obtained from this plot:
\begin{itemize}
    \item At $E_e \lesssim 1$~TeV and $B=10^{-12}$~G, the Larmor radius is $r_L < \lambda_{\rm ICS}$, meaning that deflection starts becoming important and the cascade electrons ``isotropize''. The flux calculated by $\gamma$-Cascade for point sources represents the total flux along the line of sight as well as the associated cascade halo that extends around the source\footnote{$\gamma$-Cascade does not provide information about time delay or angular spread due to magnetic fields. Such a treatment can be obtained by using Monte Carlo codes specifically designed to account for these effects \citep{Kalashev:2022cja,Blytt:2019xad}.}. Increasing the magnetic field would increase the energy at which this isotropization occurs. For unresolved source distributions, this is natural, since only a diffuse flux would be observed.

    \item At $E_e \lesssim 10^{20}$~eV,  $\lambda_{\rm ICS}$ smaller than the $\gamma$-Cascade stepsize. Therefore, many ICS interactions are expected to happen inside a single redshift bin, just as assumed in our on-the-spot approximation.

    \item Above $\sim 10^{20}$~eV, the on-the-spot approximation may no longer be valid, just as the standard cascade picture, assumed by $\gamma$-Cascade, of PP and ICS. At such high CM energies, new interactions become important, changing how the cascade develops \citep{Esmaeili:2023vyk}. In particular, electron triplet production ($e\gamma \to e e^+ e^-$) with the CMB dominates above EeV energies at $z=0$.

    \item Energy-loss through synchrotron radiation is irrelevant in electromagnetic cascades below $10^{21}$~eV for $B = 10^{-12}$~G, since its energy-loss length $\ell_{\rm sync}$ is greater than the Hubble length $d_H = c/H_0$. Assuming magnetic fields at the conservative upper limit of $\sim$~nanoGauss \citep{Planck:2015zrl}, synchrotron loss becomes significant ($\ell_{\rm sync} < \lambda_{\rm ICS}$) above a few EeV.
\end{itemize}

Synchrotron losses in $\gamma$-Cascade are parametrized through the factor 
\begin{equation}\label{eq:fICS}
    f_{\rm ICS}(E_e,z) = \frac{(dE/dt)_{\rm ICS}}{(dE/dt)_{\rm sync} + (dE/dt)_{\rm ICS}}~,
\end{equation}
which is the energy fraction going into gamma rays through ICS (while the remaining $1-f_{\rm ICS}$ is lost to synchrotron radiation). The ICS energy-loss rate is given by
\begin{equation}\label{eq:ics-E-loss-rate}
    \frac{d E_{\rm ICS}}{d t}(E_e,z) = \int dE'_\gamma \,E'_\gamma\,\gamma^{\rm ICS}_{e\to \gamma}(E_e,E'_\gamma,z)~,
\end{equation}
which is valid in the continuous-energy-loss regime, at low $E_e$. At high $E_e$, the stochastic nature of ICS becomes important, which may require a more careful treatment beyond the scope of $\gamma$-Cascade \citep{John:2022asa}. In order to account for synchrotron loss, we modify the monoenergetic on-the-spot spectra by the multiplicative factor $f_{\rm ICS}$,
\begin{equation}
    \frac{dN_{e\to \gamma,{\rm OTS}}}{dE'_\gamma}(E'_\gamma,E_e,z) \quad\to\quad f_{\rm ICS}(E_e,z)\frac{dN_{e\to \gamma,{\rm OTS}}}{dE'_\gamma}(E'_\gamma,E_e,z)~.
\end{equation}
The default (physical) IGMF is taken to be $B(z)=10^{-12}\,(1+z)^{-2}~\text{G}$ \citep{Pomakov:2022cem}, which is related to the comoving IGMF via $B(z) = (1+z)^2 B_{\rm com}(z)$ as a consequence of magnetic flux freezing.

\section{Discussion}

We now assess the performance of $\gamma$-Cascade V4 in comparison to analytical predictions of optical depths and numerical simulations from other cascade codes available in the literature.

\begin{figure}[t!]
    \centering
    \includegraphics[width=0.48\textwidth]{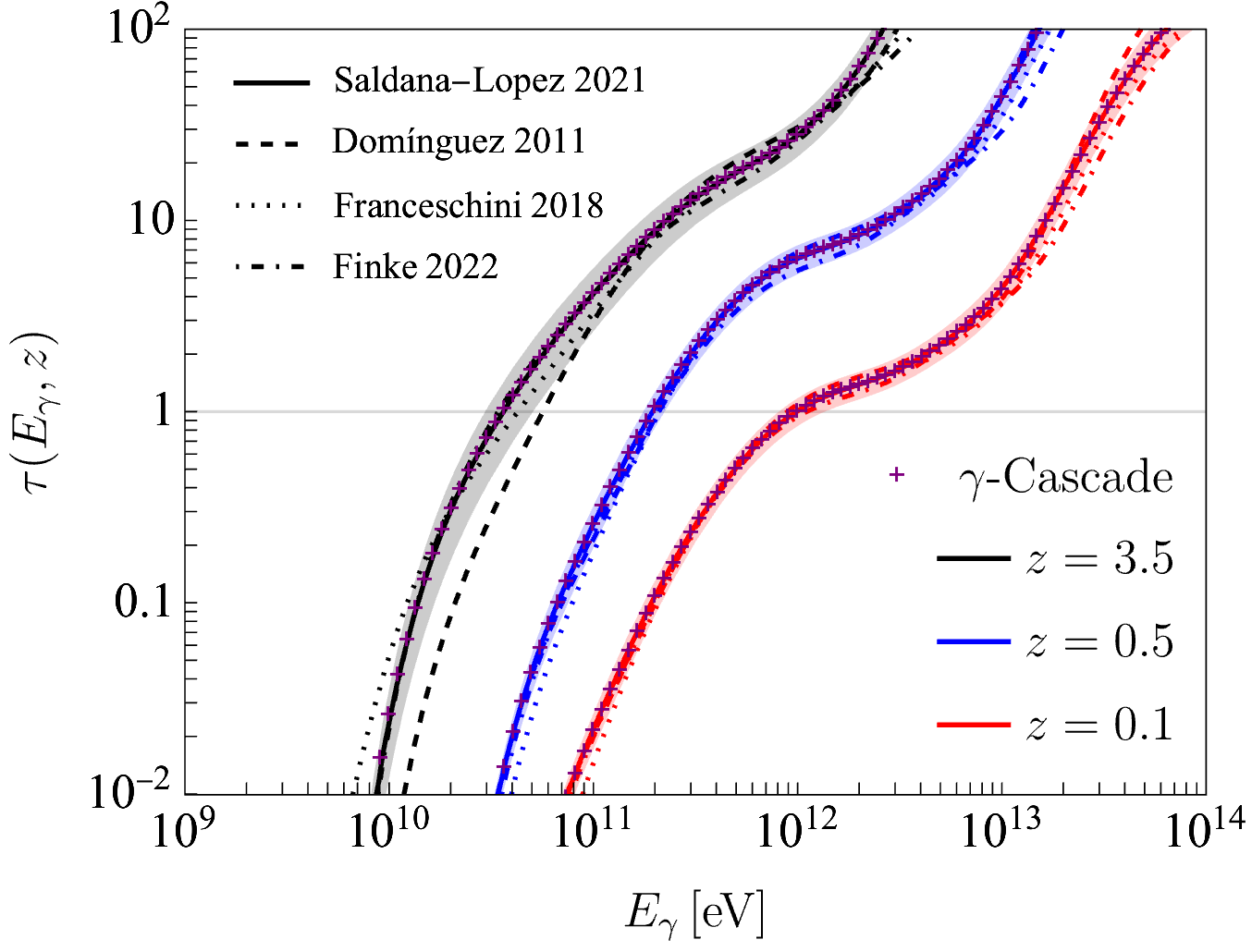}
    \caption{Optical depths for different EBL model assumptions for gamma rays emitted at redshifts of $z=0.1$, $z=0.5$, $z=3.5$ are shown in the red, blue, and black line, respectively. These were calculated using Eq. (\ref{eq:optical-depth}). $\gamma$-Cascade opacities are shown as purple crosses, and are in agreement with analytical predictions.}
    \label{fig:optical-depths}
\end{figure}

Figure \ref{fig:optical-depths} shows the optical depths, defined as follows, for gamma rays emitted at different initial redshifts $z$, predicted by all EBL models included in $\gamma$-Cascade V4, as a function of the observed gamma-ray energy at the Earth $E_\gamma$,
\begin{multline}\label{eq:optical-depth}
    \tau_{\rm PP}(E_\gamma,z) =\\ \int_{0}^{z} dz\,\frac{c}{(1+z)H(z)} \int_{m_e^2/E_\gamma(1+z)}^\infty d\epsilon\,\frac{dn(\epsilon,z)}{d\epsilon}\,\langle \sigma_{\rm PP}\rangle (E_\gamma(1+z),\epsilon)~.
\end{multline}
The shaded bands indicate the 1$\sigma$ uncertainties in the Saldana-Lopez \textit{et al.} model \citep{Saldana-Lopez:2020qzx}. To extract the optical depths directly from $\gamma$-Cascade, we evaluate the attenuated fluxes $\Phi_{\rm att}$ using the function \texttt{AttenuatePoint} (which neglects gamma-ray regeneration via ICS, see Appendix \ref{app:functions}) for each redshift, using a uniform injection spectrum\footnote{Here, $\epsilon_\gamma$ denotes the energy of gamma rays leaving the source, in contrast with $E_\gamma$ arriving at $z=0$.} $dN_\gamma/d\epsilon_\gamma = {\rm const.}$ We then take their ratios with respect to the fluxes $\Phi_{\rm redshift}$ that one would expect from cosmological redshift effects alone (without cascade interactions),
\begin{equation}
    e^{-\tau_{\rm PP}(E_\gamma,z)} = \frac{\Phi_{\rm att}(E_\gamma,z)}{\Phi_{\rm redshift}(E_\gamma,z)}~,
\end{equation}
\begin{equation}
    \Phi_{\rm redshift}(E_\gamma,z) = \frac{(1+z)^2}{4\pi d_L^2(z)}\,\frac{dN_\gamma}{d\epsilon_\gamma}(\epsilon_\gamma)\bigg|_{\epsilon_\gamma = E_\gamma(1+z)}~.
\end{equation}
Agreement is obtained between the analytical optical depths (solid lines) evaluated using Eq. (\ref{eq:optical-depth}) and the $\gamma$-Cascade simulations (purple crosses) for point sources at small, intermediate, and high redshifts. Results are identical if $\Phi_{\rm redshift}$ is evaluated numerically via the built-in function \texttt{RedshiftPoint}. This analysis serves to validate the redshifting scheme adopted by the code as well as the PP interaction rates used in internal computations.

\begin{figure}[t!]
    \centering
    \includegraphics[width=0.48\textwidth]{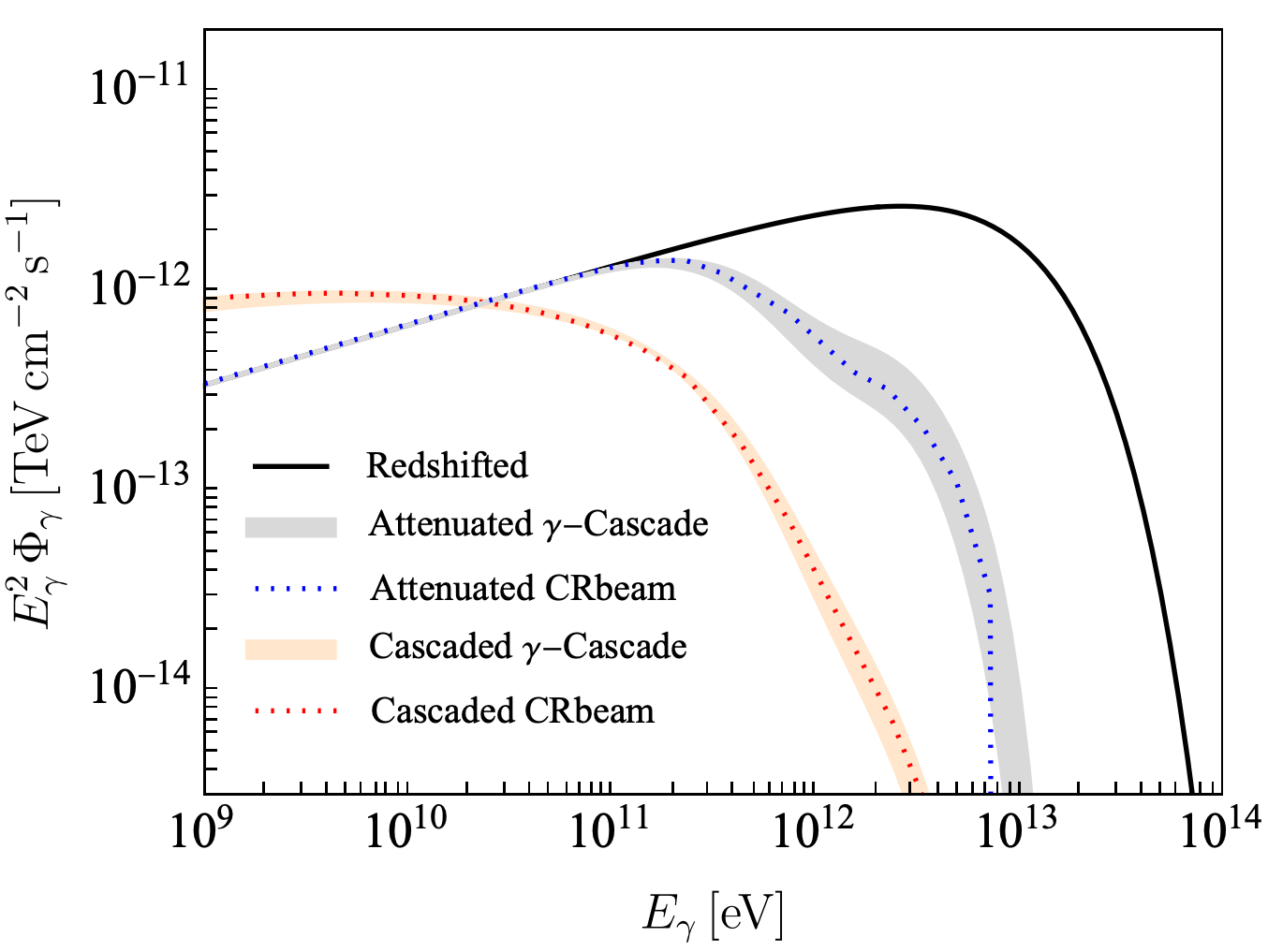}
    \caption{Comparison of $\gamma$-Cascade's attenuated (gray band) and cascaded (orange band) fluxes with CRbeam (dotted curves), for a point source at $z=0.14$ with an injection spectrum of $dN_\gamma/d\epsilon_\gamma \propto \epsilon_\gamma^{-1.7} e^{-\epsilon_\gamma/10~{\rm TeV}}$. The black solid line is the redshifted flux obtained with \texttt{RedshiftPoint}.}
    \label{fig:crbeam-comparison}
\end{figure}

Next, in Figure \ref{fig:crbeam-comparison}, we compare the outcome of $\gamma$-Cascade V4 calculations with a result from the Monte Carlo-based code CRbeam \citep{Kalashev:2022cja}. A point source at $z=0.14$ with an emission spectrum $dN_\gamma/d\epsilon_\gamma \propto \epsilon_\gamma^{-1.7}e^{-\epsilon_\gamma/10~{\rm TeV}}$ was considered, as in Figure 8 from \citep{Kalashev:2022cja}. In black we show the flux at the Earth under cosmological redshift only, while the gray band indicates the attenuated flux obtained from \texttt{AttenuatePoint} varying the Saldana-Lopez \textit{et al.} EBL model within its 1$\sigma$ uncertainty range. This region contains the attenuated flux obtained by CRbeam (blue dotted line), suggesting that EBL uncertainties dominate over the numerical differences between both codes. The orange band is the cascaded contribution to the flux reaching the Earth (once again, within the Saldana-Lopez EBL uncertainty), which is calculated by subtracting the results of \texttt{AttenuatePoint} from the corresponding results of \texttt{CascadePoint}. Once again, we find consistency with CRbeam\footnote{Since $\gamma$-Cascade is not sensitive to angular/temporal information, we reproduce CRbeam's angle- and time-integrated curve in Figure 8 labeled ``full'', under the assumption of zero magnetic field. Note that the integrated curve labeled ``jet'' is incompatible with $\gamma$-Cascade. This is because a jetted source is being considered: CRbeam accounts for the magnetic-field deviation of low-energy electrons away from the line-of-sight. An isotropic source would compensate this by electrons deviating into the line-of-sight. Although $\gamma$-Cascade cannot account for the jetted nature of the signals from such point sources, it instead calculates the cascaded flux from an isotropic emitter with the same luminosity per unit solid angle as contained in the jet.} (red dotted curve). Here again, EBL uncertainties present an important factor to take into consideration when computing cascaded fluxes at percent-level precision. We note that it was also shown in ref. \citep{Kalashev:2022cja} that CRbeam is either consistent with or improves upon other existing tools for cascade evaluation: ELMAG \citep{Blytt:2019xad} and CRPropa \citep{batista2016crpropa}. This implies consistency between $\gamma$-Cascade V4 and these codes as well.

\begin{figure}[t!]
    \centering
    \includegraphics[width=0.48\textwidth]{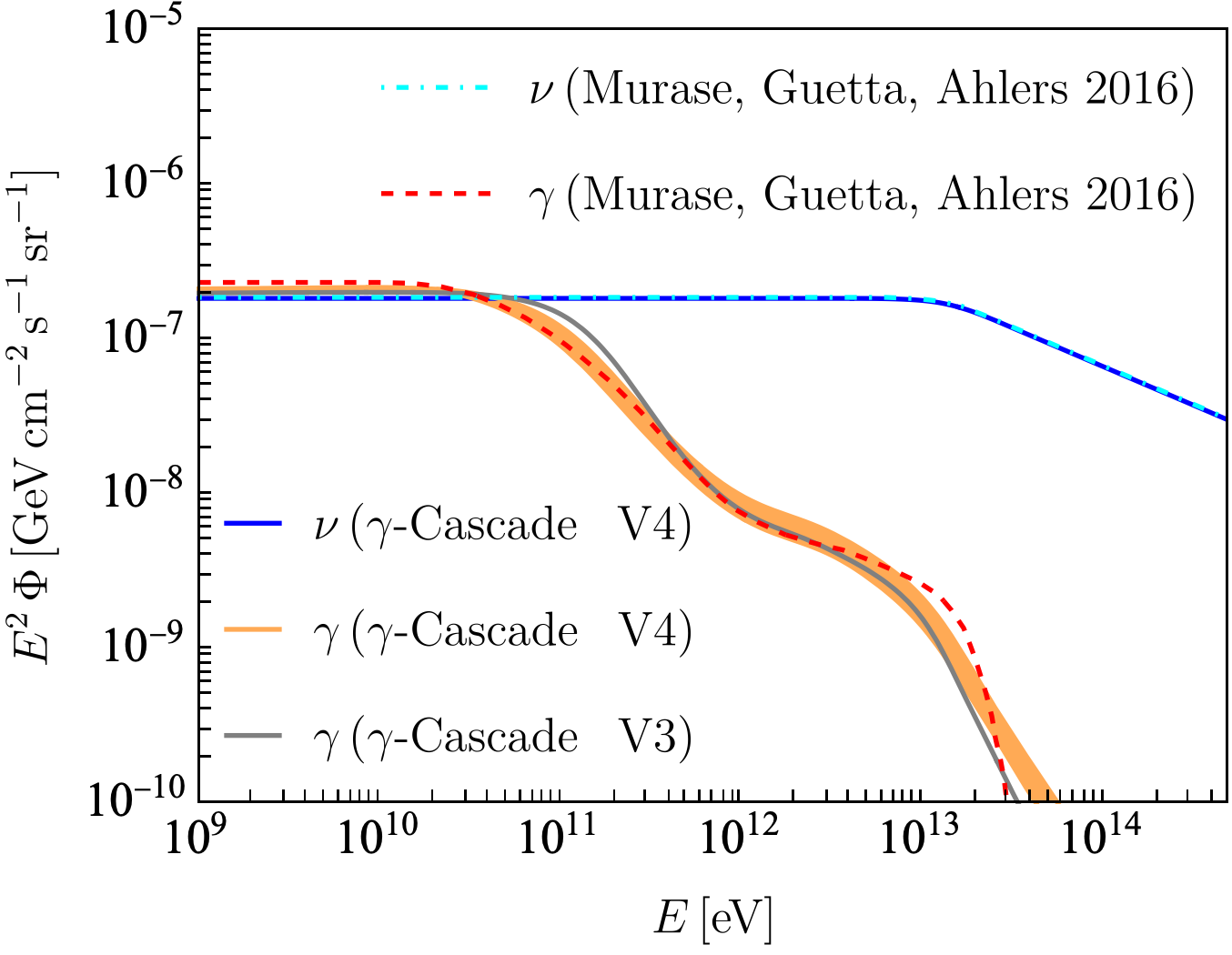}
    \caption{Diffuse cascaded fluxes obtained from $\gamma$-Cascade V4 (orange band), V3 (gray solid line), and ref.~\cite{Murase:2015xka} (red dashed line). The neutrino fluxes from the same source distribution following the model of ref.~\cite{Murase:2015xka} are also shown, with the blue curve being obtained with \texttt{RedshiftDiffuse} in $\gamma$-Cascade. See text for details of emission spectra and redshift evolution of sources.}
    \label{fig:diffuse-comparison}
\end{figure}

Finally, we compare in Figure \ref{fig:diffuse-comparison} a calculation of a cascade from a diffuse population of sources using $\gamma$-Cascade with that of \citep{Murase:2015xka}. Sources were assumed to emit neutrinos and gamma rays with broken power-law spectra given by,
\begin{equation}
    \frac{dN_\nu}{d\epsilon_\nu}(\epsilon_\nu) \propto \begin{cases}
        \epsilon_\nu^{-2} & \epsilon_\nu \leq 25~{\rm TeV}\\
        \epsilon_\nu^{-2.5} & \epsilon_\nu > 25~{\rm TeV}
    \end{cases}~,
\end{equation}
\begin{equation}
    \frac{dN_\gamma}{d\epsilon_\gamma}(\epsilon_\gamma) = \frac{1}{6}\frac{dN_\nu}{d\epsilon_\nu}(\epsilon_\nu) \bigg|_{\epsilon_\nu = \epsilon_\gamma/2}~,
\end{equation}
as in the $pp$ scenario in ref.~\citep{Murase:2015xka}, and follow the star-formation rate evolution in redshift\footnote{The assumed distribution of sources in redshift was not explicitly stated in ref. \citep{Murase:2015xka}, and neither was the EBL model considered.} \citep{Yuksel:2008cu}. The blue neutrino curve was obtained in $\gamma$-Cascade using the \texttt{RedshiftDiffuse} function (since neutrinos do not interact) and matches the cyan dot-dashed curve from \citep{Murase:2015xka}. The cascaded gamma-ray fluxes are also consistent with each other at the $\sim$1$\sigma$ level; the red dashed curve from \citep{Murase:2015xka} is almost completely contained inside the orange band, obtained by varying the Saldana-Lopez EBL model within its uncertainty range using \texttt{CascadeDiffuse} in $\gamma$-Cascade. A mild difference in the predicted cascade shape, with respect to $\gamma$-Cascade V3, can also be seen in the gray line.

\section{Conclusion}

With the increased precision of high-energy gamma-ray measurements in current and future telescopes, the necessity arises for accurate modeling of gamma-ray transport through intergalactic space. Above $\sim 100$~GeV, interactions with ubiquitous photon backgrounds lead to the formation of electromagnetic cascades. We have shown that this cascade evolution does not have a closed-form solution, and therefore requires numerical methods to accurately predict the resulting gamma-ray spectra. In this work, we have presented an updated version (V4) of the public code $\gamma$-Cascade, containing several tools for calculating cascaded gamma-ray fluxes between 0.1--$10^{12}$~GeV. This software serves to complement existing packages available in the literature, such as CRbeam and ELMAG. However, in contrast to Monte Carlo approaches, that are more suitable for point-source simulations, $\gamma$-Cascade uses a semi-analytical treatment which is ideal for obtaining diffuse fluxes from evolving populations of sources.

In this work, we detail the internal workings of $\gamma$-Cascade, with a special focus on the impact of EBL models as well as our key assumption of on-the-spot ICS spectra. In particular, we formulate the algorithm behind the cascade routine in terms of the cascade cycle shown in Figure \ref{fig:on-the-spot-scheme}. We justify the approximations made by $\gamma$-Cascade by showing that its computed spectra are in line with state-of-the-art Monte-Carlo calculations and converge onto the known universal cascade spectral shape. 

Version 4 of $\gamma$-Cascade implements several improvements over previous versions. Notably, it adds the ability to select between various different EBL models, allowing for an assessment of their impact on the resulting fluxes. Additionally, it has a more detailed treatment of spectra evaluated on the discrete energies grid, which results in a significant improvement in precision. We have shown that the accuracy of $\gamma$-Cascade is compatible with other state-of-the-art codes, finding that differences between their results are subdominant with respect to EBL uncertainties. As an example, we've shown that $\gamma$-Cascade can recreate known results.

\section*{CRediT authorship contribution statement}
\textbf{Antonio Capanema}: Conceptualization, Formal analysis, Investigation, Methodology, Software, Validation, Writing – original draft, Writing – review \& editing. \textbf{Carlos Blanco}: Conceptualization, Formal analysis, Investigation, Methodology, Software, Validation, Writing – original draft, Writing – review \& editing.

\section*{Declaration of competing interest}
The authors declare that they have no known competing financial interests or personal relationships that could have appeared to influence the work reported in this paper.

\section*{Data availability}
The full $\gamma$-Cascade V4 package, including the main code along with its essential and auxiliary file libraries, has been made available for download at \url{https://github.com/GammaCascade/GCascade}.

\section*{Acknowledgements}
Special thanks to Arman Esmaili and to Markus Ahlers, for useful discussions and suggestions. A.C. thanks for the support received by the scholarships CAPES/PROEX No. 88887.511843/2020-00, CNPq No. 140316/2021-3, CAPES-PrInt No. 88887.717489/2022-00 and FAPERJ No. E-26/204.138/2022. The work of C.B.~was supported in part by NASA through the NASA Hubble Fellowship Program grant HST-HF2-51451.001-A awarded by the Space Telescope Science Institute, which is operated by the Association of Universities for Research in Astronomy, Inc., for NASA, under contract NAS5-26555.

\appendix
\section{Energy Conservation on a Grid}
\label{app:grid}

Evaluating PP and ICS spectra for initial and final energies constrained to the \texttt{energies} grid can lead to a few issues. More precisely, after calculating the spectra given in Eqs. (\ref{eq:ppspec-monoenergetic}) and (\ref{eq:ics-spec-monoenergetic}) at discrete energy values, one must be careful to guarantee that Eqs. (\ref{eq:pp-number-cons}), (\ref{eq:pp-energy-cons}), (\ref{eq:ics-number-cons}) and (\ref{eq:ics-energy-cons}) are respected. If not, energy may leak away from or into the cascade, leading to unphysical results. There are several features of these spectra that cause such problems. In general, these numerical problems are solved by numerical integrators by employing adaptive sampling techniques. However, since we know the behavior of these functions a priori, we can perform numerical post processing in order to avoid using time-consuming adaptive algorithms. To illustrate these numerical pitfalls, we consider some concrete examples below and discuss how to fix them without compromising the physics.

\begin{figure}[t!]
    \centering
    \includegraphics[width=0.48\textwidth]{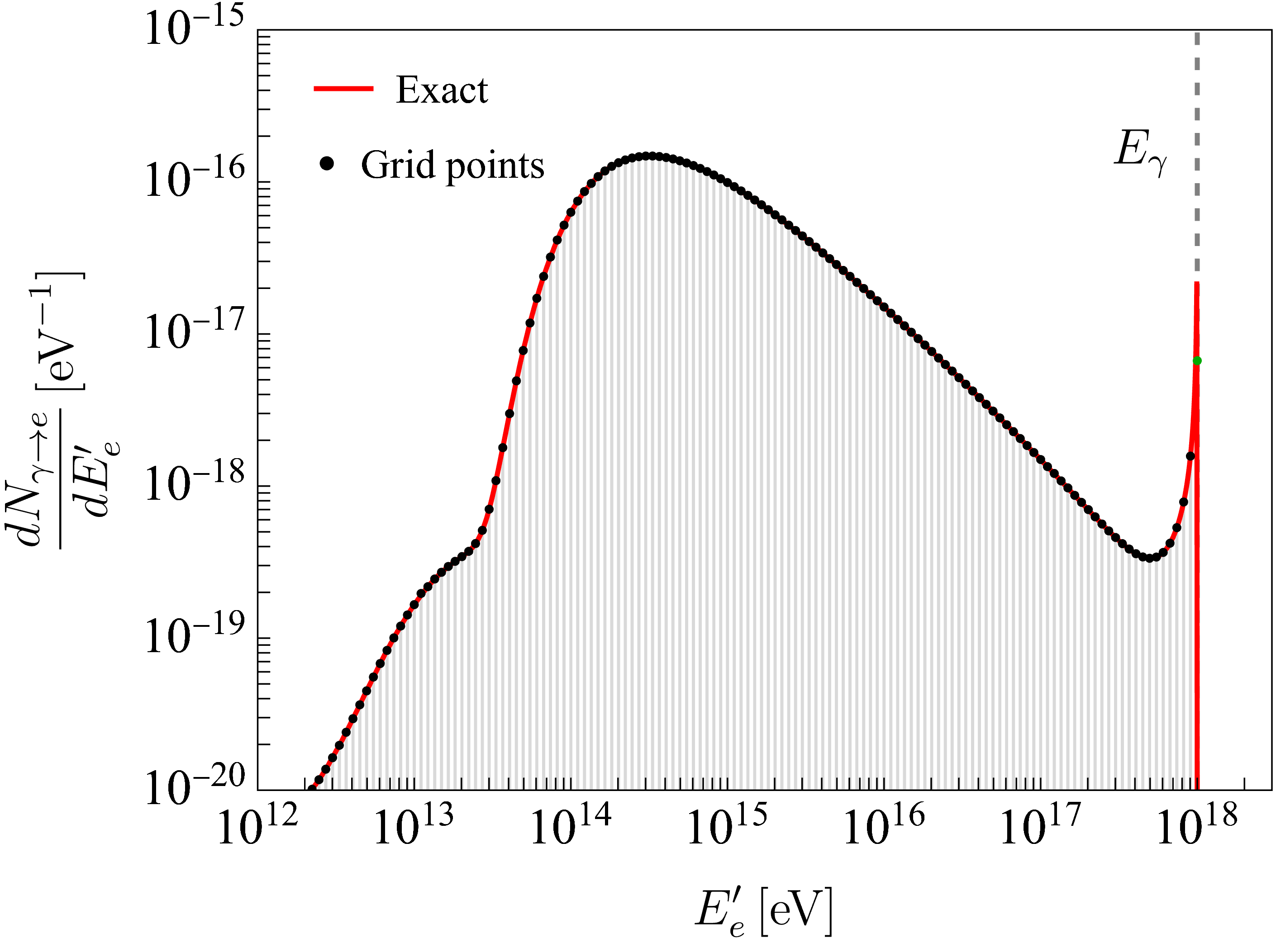}
    \caption{Electron spectrum from a monoenergetic flux of $E_\gamma = 1$~EeV gamma rays at $z=0$. The exact spectrum is given by the red curve, while the black points mark its values at the energies in the \texttt{energies} grid.}
    \label{fig:grid1}
\end{figure}

\begin{figure}[t!]
    \centering
    \includegraphics[width=0.48\textwidth]{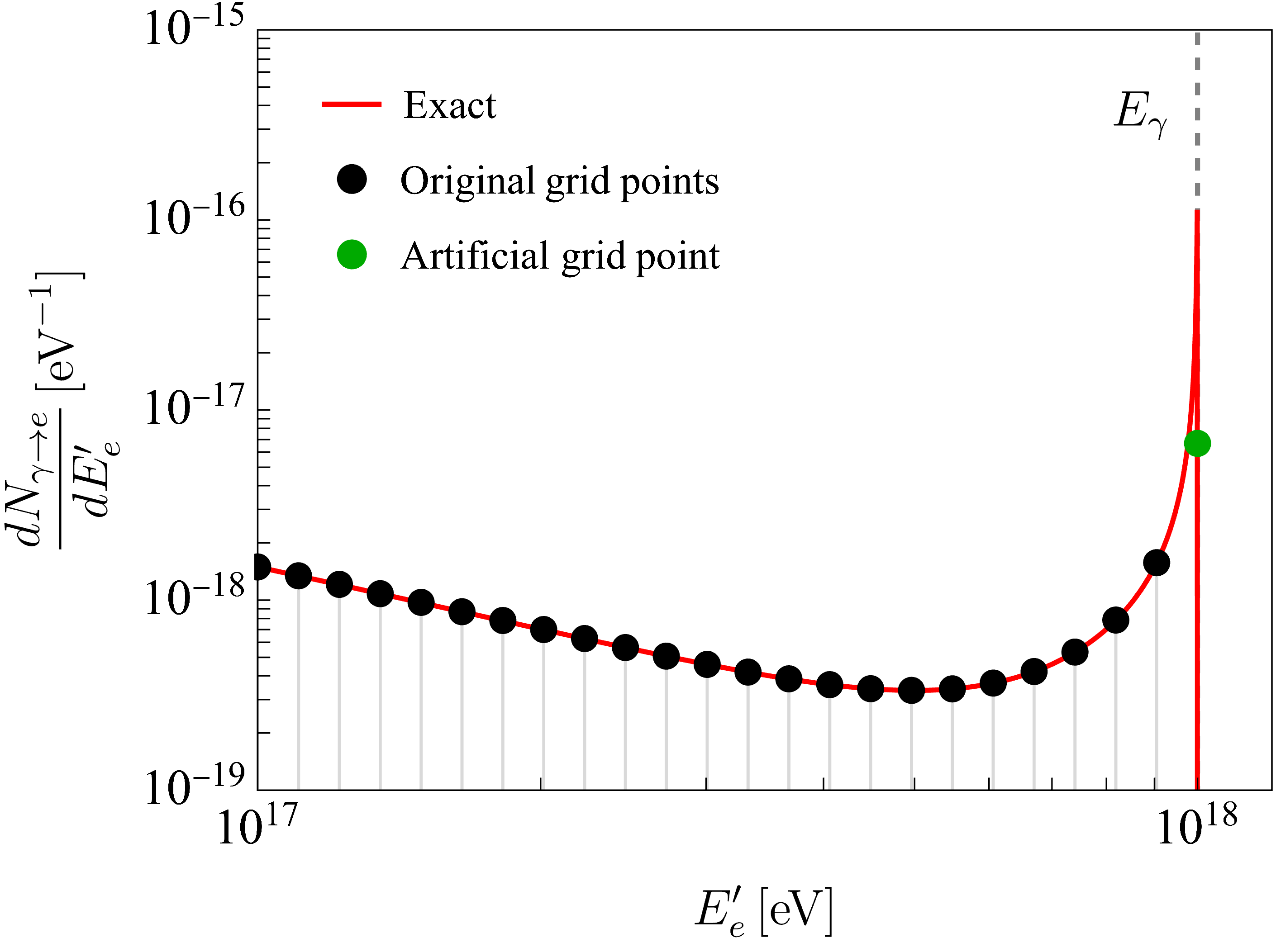}
    \caption{Zoom into the region around 1~EeV, showing the artificial grid point (in green) added to enforce energy/particle number conservation.}
    \label{fig:grid1zoom}
\end{figure}

First, let us examine the issues arising in PP spectra. Consider a monoenergetic flux of gamma rays with energy $E_\gamma = 1$~EeV at redshift $z=0$, all of which interact with the CMB/EBL to generate an electron spectrum given by Eq. (\ref{eq:ppspec-monoenergetic}). Figure \ref{fig:grid1} shows the exact spectrum in red, overlapped by black points along the \texttt{energies} grid at which the spectrum was evaluated. Notice how the spectrum is highly peaked  at nearly 1 EeV (i.e. the width of the peak is smaller than the width of the energy grid spacing $\Delta E_e$), which is expected from the low inelasticity of PP at such a high CM energy. This region is where most of the outgoing energy is concentrated. Meanwhile, at precisely 1 EeV it evaluates exactly to zero, which is why there is no grid point along the vertical gray dashed line marked by $E_\gamma$, even though 1 EeV is a value in the \texttt{energies} grid. As a result, the trapezoidal integration over the grid points given in Eq. (\ref{eq:pp-number-cons-trapint}) will indicate missing energy/particle number in the outgoing spectrum,
\begin{multline}\label{eq:trapint}
    \int dE'_e \,E'_e\,\frac{d N_{\gamma\to e}}{d E'_e}(E'_e,E_\gamma,z) \\
    \approx \sum_{k=1}^{299} \left[E_{k+1}\left(\frac{dN_{\gamma\to e}}{dE'_e}\right)_{i,j,k+1} + E_k\left(\frac{dN_{\gamma\to e}}{dE'_e}\right)_{i,j,k} \right]\frac{E_{k+1} - E_k}{2} < E_\gamma~.
\end{multline}
To fix this, we introduce a non-zero value for the spectrum at the 1 EeV grid point, representing the energy in electrons of peak energy $E_\gamma - \delta E_e$ for $\delta E_e \ll \Delta E_e$, in order to enforce Eq. (\ref{eq:pp-energy-cons}), and consequently (\ref{eq:pp-number-cons}), under trapezoidal integration. The result of this procedure is shown in Figure \ref{fig:grid1zoom}, zooming in to the region around 1 EeV, where the green point was introduced to compensate the energy deficit. Although this solution leads to the unphysical creation of electrons at $E_\gamma$, it is justified by the peaked nature of the spectrum, and reproduces all the important features required for a reliable electromagnetic cascade simulation.

\begin{figure}[t!]
    \centering
    \includegraphics[width=0.48\textwidth]{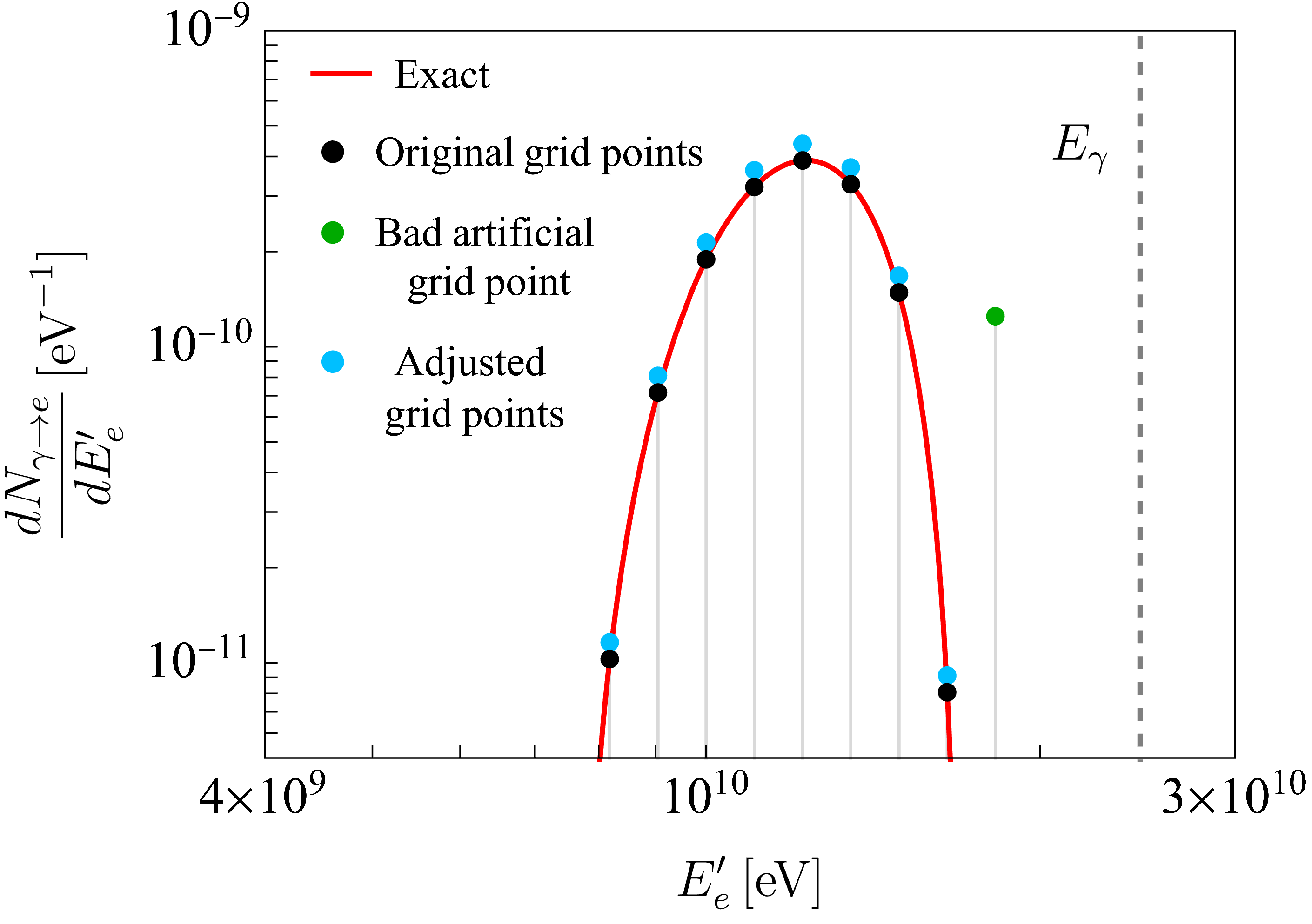}
    \caption{Same as Figure \ref{fig:grid1}, but for $E_\gamma \approx 25$~GeV. Although adding the green point fixes energy conservation, it does not reflect the behavior of the spectrum. Instead, the black grid points are all raised by a small constant value, becoming the blue points.}
    \label{fig:grid2}
\end{figure}

In some cases, where the spectra drop rapidly within a grid width, the energy deficit is too large, causing the previous fix to  introduce a point that is unreasonably high when compared to the spectrum. Figure \ref{fig:grid2} displays this situation for a PP spectrum from gamma rays at $\texttt{energies[56]} \approx 25$~GeV, where the green artificial point is clearly too high. When this happens, instead of employing the previous solution (which is only appropriate for peaked spectra), we compensate the missing energy/particles by multiplying the entire grid by a constant factor (which is only slightly greater than 1), as shown by the blue points in Figure \ref{fig:grid2}.

\begin{figure}[t!]
    \centering
    \includegraphics[width=0.48\textwidth]{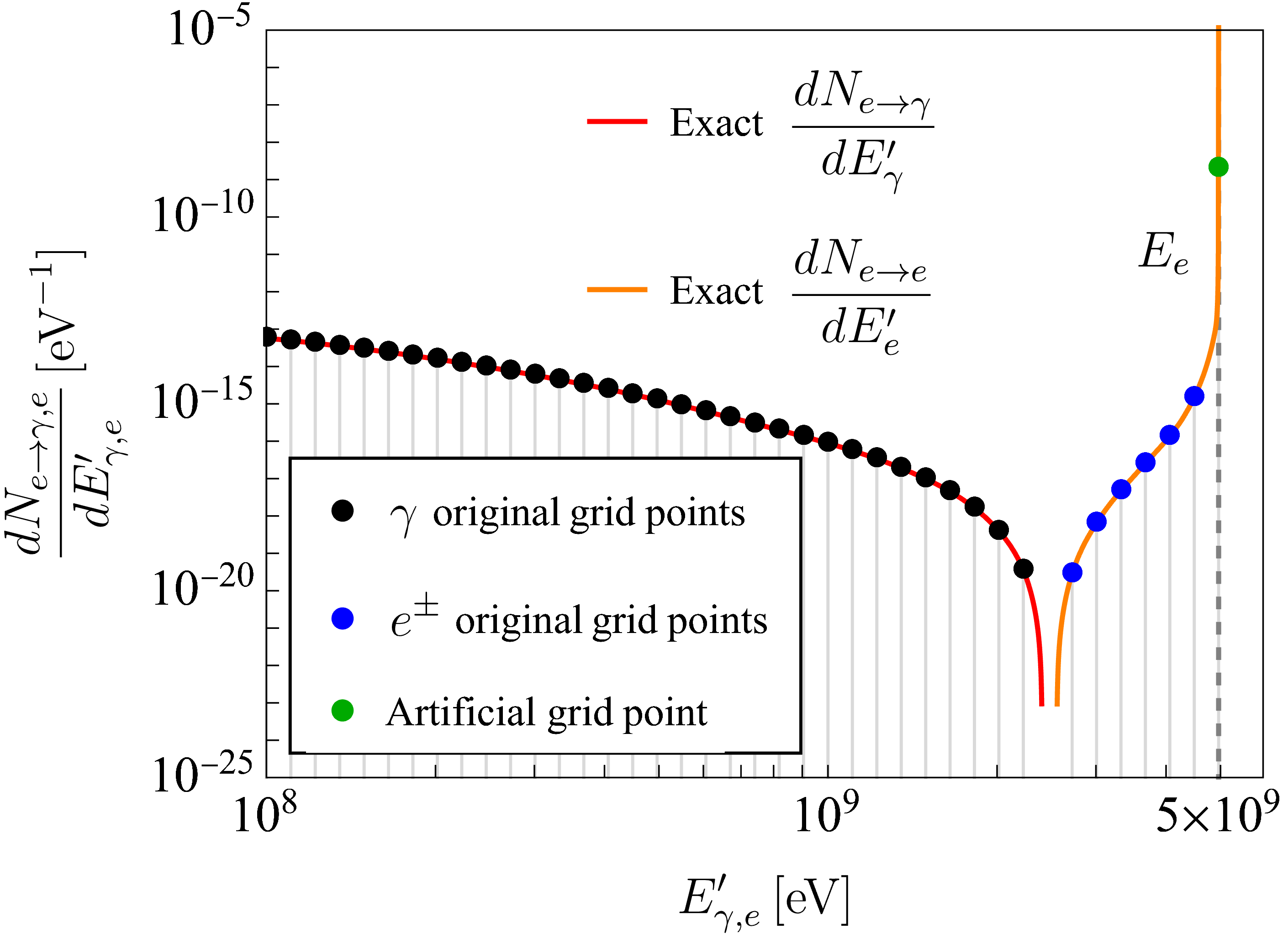}
    \caption{Gamma-ray and electron spectra from a monoenergetic flux of $E_e \approx 5$~GeV electrons at $z=0$. Black and blue grid points mark the values of the gamma-ray and $e^\pm$ spectra, respectively, at the \texttt{energies} grid values. To fix the electron grid, we artificially add the green point such that particle number/energy conservation is preserved. Note that a significant portion of the gamma-ray spectrum lies below $10^8$~eV. We discuss the enforcement of conservation beyond the limits of the \texttt{energies} grid in Appendix \ref{app:on-the-spot}.}
    \label{fig:grid3}
\end{figure}

A third an final problem that arises in PP spectra occurs when a steeply falling spectrum causes the trapezoidal sum in Eq. (\ref{eq:trapint}) to be slightly $>E_\gamma$. Employing the fix from the first scenario would generate an adjusted point with a negative value. In this case, we enforce Eq. (\ref{eq:pp-energy-cons}) by once again artificially multiplying the entire grid by a constant factor (this time, slightly smaller than 1).

In the case of ICS, grid-related problems can arise in both the outgoing gamma-ray and/or electron spectra. Luckily, their spectral shapes exhibit similar features to the PP spectra, meaning we can fix both the gamma-ray and electron ICS grids independently by employing the same three kinds of modifications described above. However, the behavior of ICS at low energies leads to an additional issue: the gamma-ray spectrum extends itself well below $10^8$~eV, where the \texttt{energies} array ends. This is depicted in Figure \ref{fig:grid3}, which shows the outgoing photon and electron spectra from the ICS of $E_e = \texttt{energies[40]} \approx 5$~GeV electrons. The black and blue points mark the values from their corresponding $d N_{e\to \gamma}/ d E'_\gamma$ and $d N_{e\to e}/ d E'_e$ grids, respectively, while the green point was adjusted in the electron grid to conserve particle number and mimic the exact spectrum's behavior. Clearly, trying to enforce particle number conservation on the gamma-ray spectrum by trapezoidal integration along the \texttt{energies} grid is not appropriate. It is also not necessary, as $\gamma$-Cascade should really lose part of its initial particle number/energy to the $<~10^8$~eV region. Since these are well-normalized by design, we leave that grid untouched for now.

\section{On-The-Spot Approximation on a Grid}
\label{app:on-the-spot}

The on-the-spot approximation requires the evaluation of several generations of subsequent ICS interactions, as depicted in Figure \ref{fig:on-the-spot-scheme}. An important issue arises when performing these calculations on a static energy grid. As $E_e$ decreases, the outgoing electron spectra from ICS become sharply peaked and constrained to a narrow energy region at just below $E_e$. This physically corresponds to the dwindling energy loss. If this is allowed to proceed unchecked, the ICS loss stalls completely, leading to an unphysical numerical pileup of gamma rays at low energies. For our on-the-spot approximation to work on a grid, the outgoing ICS spectra $dN_{e\to \gamma,e}/dE_{\gamma,e}$ from Eq. (\ref{eq:ics-spec-monoenergetic}) are adjusted beyond the post-processing already applied for energy conservation in Appendix \ref{app:grid}. This is done in the following procedure: 
\begin{enumerate}
    \item All post-processing described in Appendix \ref{app:grid} are applied to the ICS spectra.
    
    \item Check if ICS energy loss is stalling, which occurs when the ICS energy loss in a given cascade generation falls below the resolution of the energy grid. This happens when following condition is satisfied, 
    \begin{equation}
        \int dE'_\gamma \,E'_\gamma\,\frac{dN_{e\to \gamma}}{dE'_\gamma}(E'_\gamma,E_e,z) <  \Delta E,
    \end{equation}
    where it's understood that the integrals are discrete (i.e. trapezoidal integration), and $E_e - \Delta E$ is the energy grid point directly below $E_e$ in the \texttt{energies} array.
    
    \item If stalled, modify the spectrum of outgoing electrons from an initial energy $E_e$, such that the new spectra evaluated at the energy grid are just as if the initial electron's energy had been $E_e - \Delta E$,
    \begin{equation}\label{eq:modified-electron-grid}
        \frac{dN_{e\to e}}{dE'_e}(E'_e,E_e,z) \to \frac{dN_{e\to e}}{dE'_e}(E'_e,E_e - \Delta E,z)~.
    \end{equation}
    This forces ICS to progress from $E_e$ to $E_e - \Delta E$ in a single generation, instead of over many generations. Physically, this corresponds to enforcing that the energy loss from ICS is proportional to the energy of the electron.
    
    \item The energy difference is deposited into ICS gamma rays. These gamma rays are produced following the ICS spectral distribution from an electron at $E_e$ renormalized such that energy conservation (\ref{eq:ics-energy-cons}) is satisfied\footnote{This requires extending the \texttt{energies} array to lower energies to get its total energy, since a considerable fraction of the spectrum is below 0.1 GeV.},
    \begin{multline}\label{eq:rescaling-gamma-grid}
        \frac{dN_{e\to \gamma}}{dE'_\gamma}(E'_\gamma, E_e, z) \\ \to \frac{dN_{e\to \gamma}}{dE'_\gamma}(E'_\gamma, E_e, z) \times \frac{\Delta E}{\int dE'_\gamma \,E'_\gamma\,\frac{dN_{e\to \gamma}}{dE'_\gamma}(E'_\gamma,E_e,z)}~.
    \end{multline}
\end{enumerate}
This last step comes at the cost of abandoning particle number conservation for gamma rays in a given generation, but this is justified. Electrons at $E_e$ slowly lose their energy through many ICS interactions and eventually reach $E_e - \Delta E$ within a Hubble timescale. Since we are essentially condensing many interactions into a single one, this naturally increases the outgoing particle number.

\begin{figure}[t!]
    \centering
    \includegraphics[width=0.48\textwidth]{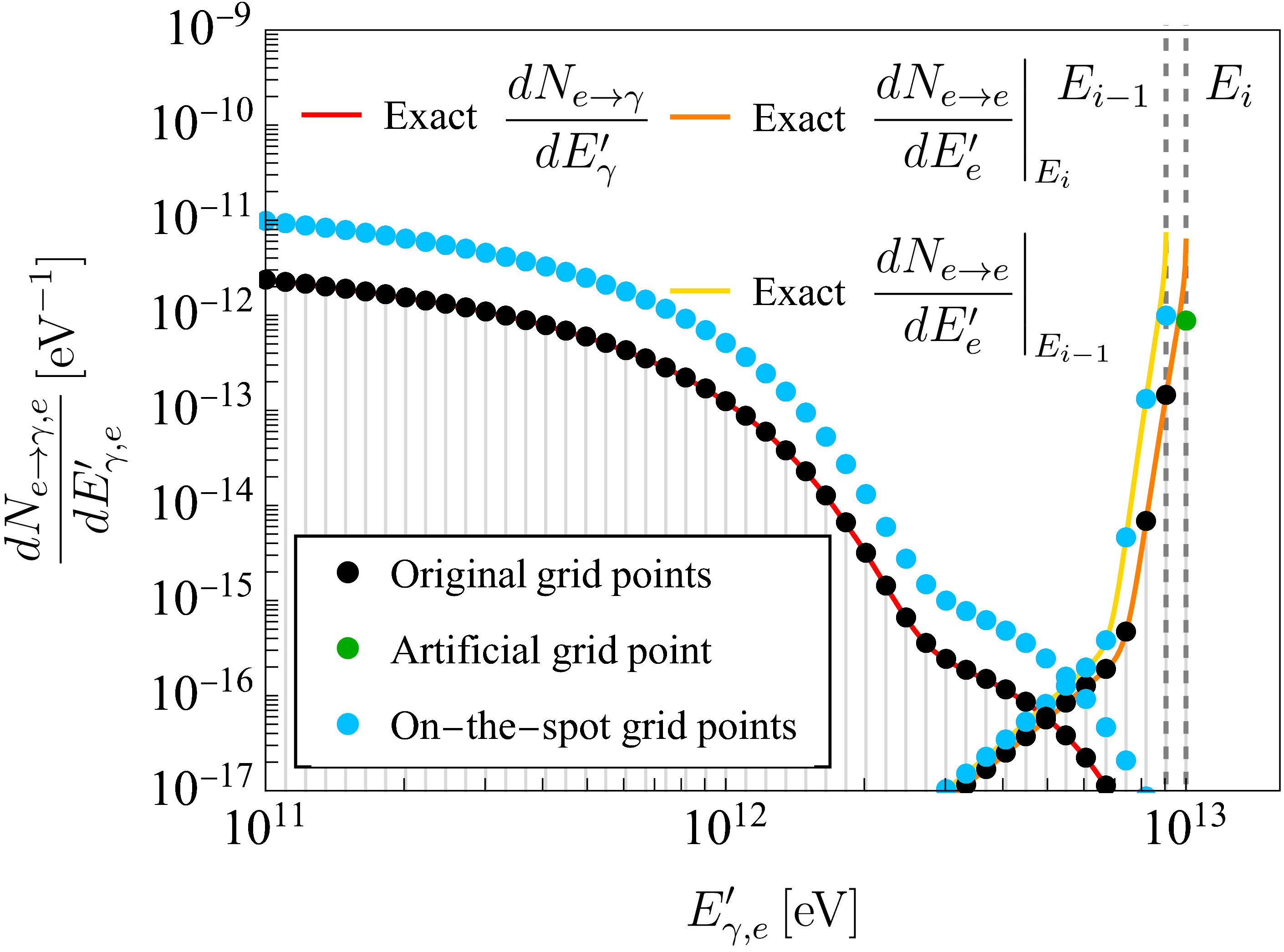}
    \caption{Points evaluated on the energy grid for ICS spectra from monoenergetic electrons before any post processing (black points), after ensuring energy conservation (added green point; see Appendix \ref{app:grid}), and after the modifications required for our on-the-spot approximation (blue points). See text for the description and justification behind these modifications. The exact spectra were obtained for initial electron energies of $E_i =$~\texttt{energies[116]}~$=10$~TeV and $E_{i-1} =$~\texttt{energies[115]}. Note that the modified electron points for $E_i$ follow the $E_{i-1}$ spectrum, avoiding the stalling problem of electrons at $E_i$.}
    \label{fig:ots-grid}
\end{figure}

The result of this procedure is shown in Figure \ref{fig:ots-grid}, where we display the outgoing spectra, along with the energy-grid evaluated points associated to them, from the ICS of a monoenergetic flux of electrons with $E_e = 10$~TeV (labeled by $E_i$) at $z=0$. Black points are obtained evaluating the exact spectra $dN_{e\to \gamma,e}/dE'_{\gamma,e}$ over the \texttt{energies} array. The green point in is the result of post processing in order to enforce energy conservation, as described in Appendix \ref{app:grid}. Since this point represents essentially all of the initial energy, the electrons are effectively stuck at $E_e$, i.e. the ICS energy loss has stalled. We further adjust the outgoing electron and gamma-ray spectra such that they become the blue points. The new gamma-ray spectrum is a rescaling of its original shape, shown as the  black points, and given by Eq. (\ref{eq:rescaling-gamma-grid}). The new electron spectrum is equal to the one obtained from initial monochromatic electrons at the next lower energy in the \texttt{energies} array, $E_e - \Delta E$ (labeled by $E_{i-1}$), corrected for energy conservation, as in Eq. (\ref{eq:modified-electron-grid}). Thus, ICS progresses to lower energies with each generation.

\begin{figure}[t!]
    \centering
    \includegraphics[width=0.48\textwidth]{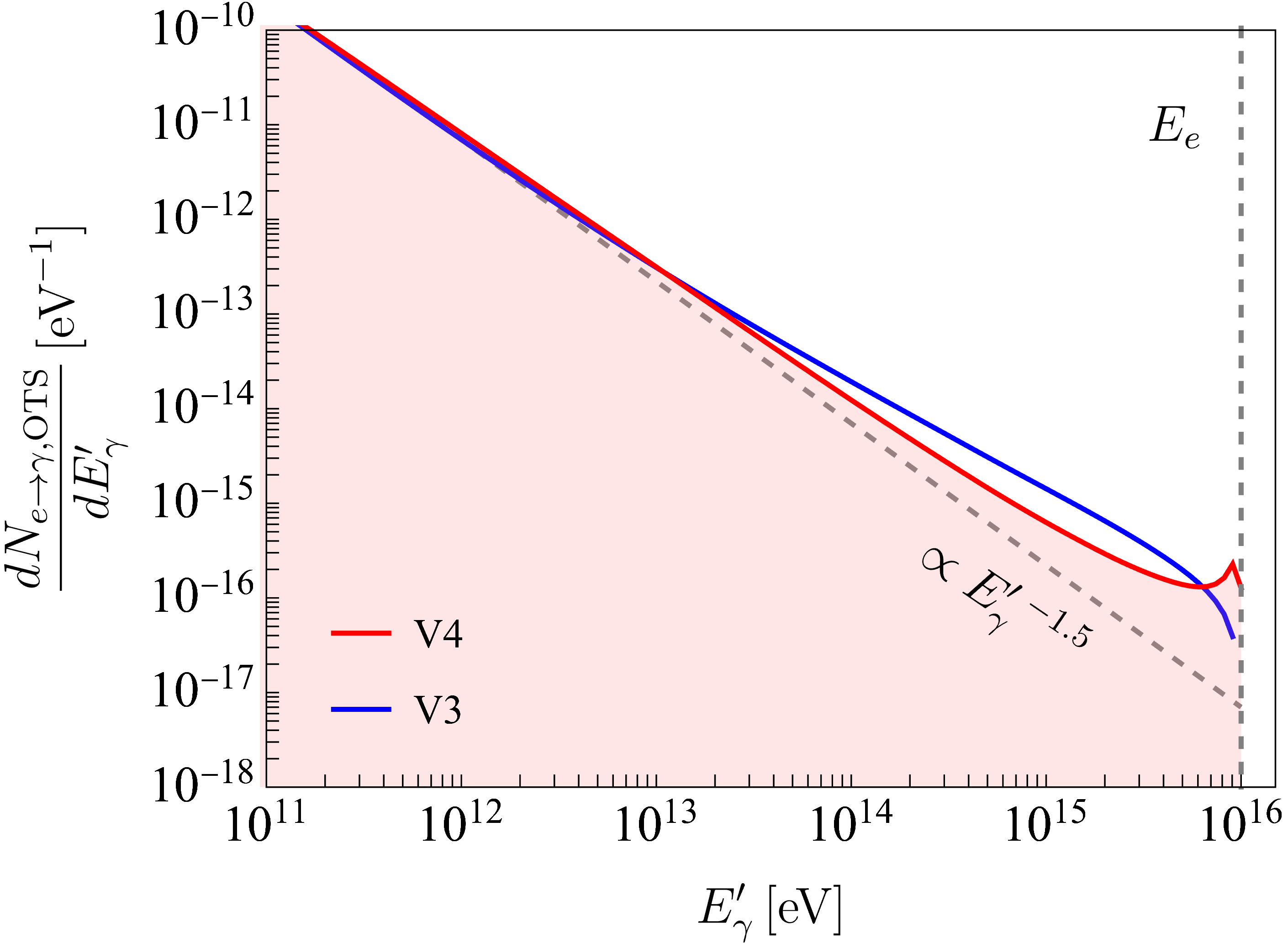}
    \caption{On-the-spot gamma-ray spectrum from electrons initially at $E_e = 10$~PeV. The red and blue curves are the current (V4) and previous (V3) version's results interpolated over the energy grid, respectively. Both converge to Berezinsky's well-known analytical prediction $dN/dE'_\gamma \propto E_\gamma^{\prime\,-1.5}$ at low energies \citep{Berezinsky:2016feh}.}
    \label{fig:ots-grid-final}
\end{figure}

Having adjusted the spectra evaluated on our grid, we obtain the corresponding values for $dN_{e\to \gamma,{\rm OTS}}/dE'_\gamma$ iteratively through the process described in Eqs. (\ref{eq:on-the-spot-spec})--(\ref{eq:on-the-spot-nth-espec}). Figure \ref{fig:ots-grid-final} shows the resulting on-the-spot spectrum, in red, for initial monoenergetic electrons at $E_e = 10$~PeV and $z=0$, interpolated over the energy grid. For comparison, we show in blue the same spectra as implemented in $\gamma$-Cascade V3. The corrections at high energies are due to the energy conservation-enforcing post-processing described in Appendix \ref{app:grid}. Note that the spectrum correctly follows the universal $dN/dE'_\gamma \propto dE_\gamma^{\prime\,-1.5}$ analytical prediction from \citep{Berezinsky:2016feh} for cascades in the low energy regime (where only ICS occurs).

\section{$\gamma$-Cascade Functions}
\label{app:functions}

$\gamma$-Cascade provides nine built-in gamma-ray propagation functions: \texttt{RedshiftPoint}, \texttt{AttenuatePoint}, \texttt{CascadePoint}, \texttt{RedshiftDiffuse}, \texttt{AttenuateDiffuse}, \texttt{CascadeDiffuse}, \texttt{RedshiftEvolving}, \texttt{AttenuateEvolving} and \texttt{CascadeEvolving}, along with two additional functions that allow the user to change the EBL model and the IGMF strength\footnote{$\gamma$-Cascade is not sensitive to its coherence length; only synchrotron losses are taken into account, not magnetic deflection.} adopted by the code: \texttt{changeEBLModel} and \texttt{changeMagneticField}, respectively. By default, $\gamma$-Cascade considers the EBL model by Saldana-Lopez \textit{et al.} \citep{Saldana-Lopez:2020qzx} and a physical magnetic field of $B(z)=10^{-12}(1+z)^{-2}$~G. Here we present all functions found in $\gamma$-Cascade; for details regarding their usage, including units of their inputs/outputs, the reader is referred to the \texttt{Tutorial.nb} notebook that comes with the $\gamma$-Cascade package. 
\begin{itemize}
\item \texttt{RedshiftPoint[injected spectrum, source redshift]} produces the redshifted flux at Earth ($z=0$) from a point source, in the absence of any interactions.

\item \texttt{AttenuatePoint[injected spectrum, source redshift]} produces the attenuated flux at Earth from a point source, without including cascade evolution (\textit{i.e.} accounting only for PP attenuation and cosmological energy redshifting, without the regeneration of gamma rays via ICS).

\item \texttt{CascadePoint[injected spectrum, source redshift]} produces the inclusive flux at Earth from a point source, accounting for electromagnetic cascades (\textit{i.e.} taking into account PP, ICS, synchrotron losses and cosmological redshifting).

\item \texttt{RedshiftDiffuse[injected spectrum, maximum source redshift, comoving density distribution of sources]} produces the diffuse flux at Earth from a population of identical sources (\textit{i.e.} with the same emission spectra) following a given density distribution per comoving volume as a function of redshift, taking into account cosmological redshifting only (no interactions).

\item \texttt{AttenuateDiffuse[injected spectrum, maximum source redshift, comoving density distribution of sources]} produces the diffuse flux at Earth from a population of identical sources following a given density distribution per comoving volume as a function of redshift, taking into account PP attenuation and redshifting effects, without gamma-ray regeneration via ICS.

\item \texttt{CascadeDiffuse[injected spectrum, maximum source redshift, comoving density distribution of sources]} produces the diffuse flux at Earth from a population of identical sources following a given density distribution per comoving volume as a function of redshift, taking into account cosmological expansion and electromagnetic cascade evolution.

\item \texttt{RedshiftEvolving[injected spectrum, maximum source redshift, comoving density distribution of sources]} produces the diffuse flux at Earth from a population of evolving sources (\textit{i.e.} with redshift-dependent intrinsic spectra) following a given comoving density distribution in redshift, taking into account cosmological redshifting only (no interactions).

\item \texttt{AttenuateEvolving[injected spectrum, maximum source redshift, comoving density distribution of sources]} produces the flux at Earth from a population of evolving sources (\textit{i.e.} with redshift-dependent intrinsic spectra) following a given comoving density distribution in redshift, taking into account PP attenuation and redshifting effects, without gamma-ray regeneration via ICS.

\item \texttt{CascadeEvolving[injected spectrum, maximum source redshift, comoving density distribution of sources]} produces the flux at Earth from a population of evolving sources (\textit{i.e.} with redshift-dependent intrinsic spectra) following a given comoving density distribution in redshift, taking into account cosmological expansion and electromagnetic cascade evolution.

\item \texttt{changeEBLModel[EBL model index]} changes the EBL model used in $\gamma$-Cascade. Each model is associated with an integer index, from \texttt{0} to \texttt{6}, which are:\\
\texttt{0} - no EBL, cascades develop under the CMB only;\\
\texttt{1} - best-fit Saldana-Lopez \textit{et al.} (2021) model \cite{Saldana-Lopez:2020qzx} (default value);\\
\texttt{2} - $1\sigma$ upper limit of the Saldana-Lopez \textit{et al.} (2021) model \cite{Saldana-Lopez:2020qzx};\\
\texttt{3} - $1\sigma$ lower limit of the Saldana-Lopez \textit{et al.} (2021) model \cite{Saldana-Lopez:2020qzx};\\
\texttt{4} - Finke \textit{et al.} (2022) model \cite{Finke:2022uvv};\\
\texttt{5} - Franceschini and Rodighiero (2017) model \cite{Franceschini:2017iwq};\\
\texttt{6} - Domínguez \textit{et al.} (2011) model \cite{Dominguez:2010bv} (used in $\gamma$-Cascade V3).

\item \texttt{changeMagneticField[field strength, redshift evolution index, EBL index]} changes the IGMF for the chosen EBL model, producing new library files for the given field strength $B_0$ and redshift evolution spectral index $\gamma$, following $B(z)=B_0\,(1+z)^{-\gamma}$. The previous library files have their names prepended with ``\texttt{bak\char`_}''.
\end{itemize}

Computational time is approximately linear in max distance, $z$, for all functions listed above. A reference computer running on 32Gb of RAM and an AMD Ryzen 7040 series CPU (16 physical cores 3.2GHz up to 5.3 GHz) was used to characterize the computational time using a flat power spectrum across the entire energy range. The propagation functions were benchmarked for execution time by executing them from a max distance of $z = 0.1$ . The redshift-only propagation functions (\texttt{RedshiftPoint}, \texttt{RedshiftDiffuse}, \texttt{RedshiftEvolving}) run in approximately 0.05~s, the propagation functions that include attenuation (\texttt{AttenuatePoint}, \texttt{AttenuateDiffuse}, \texttt{AttenuateEvolving}) run in approximately 0.5~s, and the propagation functions that include cascade development (\texttt{CascadePoint}, \texttt{CascadeDiffuse}, \texttt{CascadeEvolving}) run in approximately 40~s. It should be noted that the shape of the injected spectrum does not impact computational time significantly.


\bibliographystyle{elsarticle-num-names} 
\bibliography{V4-CPC/elsarticle-template-num-names}





\end{document}